\documentclass[12pt]{article}

\usepackage{graphicx,amssymb,url,mathrsfs, amsmath}
\usepackage{eucal}
\usepackage{wrapfig}
\usepackage{boxedminipage}
\usepackage{setspace}
\usepackage{subfigure}

\def\a  {\alpha}       \def\b  {\beta}         \def\g  {\gamma}
       \def\d  {\delta}        \def\D  {\Delta}
        \def\k  {\kappa}
\def\l  {\lambda}             \def\m  {\mu}
\def\n  {\nu}                  
\def\t  {\tau}                 
   \def\w  {\omega}

\newcommand{\cala}{\mbox{${\cal A}$}}

 \newcommand{\call}{\mbox{${\cal L}$}}
\newcommand{\calm}{\mbox{${\cal M}$}} 
\newcommand{\calo}{\mbox{${\cal O}$}}

\def\IR{{\hbox{{\rm I}\kern-.2em\hbox{\rm R}}}}
\def\IB{{\hbox{{\rm I}\kern-.2em\hbox{\rm B}}}}
\def\IN{{\hbox{{\rm I}\kern-.2em\hbox{\rm N}}}}
\def\IC{\,\,{\hbox{{\rm I}\kern-.59em\hbox{\bf C}}}}
\def\IZ{{\hbox{{\rm Z}\kern-.4em\hbox{\rm Z}}}}
\def\IP{{\hbox{{\rm I}\kern-.2em\hbox{\rm P}}}}
\def\IH{{\hbox{{\rm I}\kern-.4em\hbox{\rm H}}}}
\def\ID{{\hbox{{\rm I}\kern-.2em\hbox{\rm D}}}}

\def\be{\begin{equation}}
\def\ee{\end{equation}}
\def\ba{\begin{eqnarray}}
\def\ea{\end{eqnarray}}

\def\half{\frac{1}{2}}

\newcommand{\inv}[1]{\frac{1}{#1}}

\def\ra{\rightarrow}  

\newcommand{\dint}[2]{\int_{#1}^{#2}\!\!}

\newcommand{\del}[1]{\partial_{#1}}
\def\dell{\partial}

\def\Tr{{\rm tr}\,}

\def\det{{\rm det}}

\def\nn{\nonumber}
\def\ea{{\it et al}. }

\newcommand{\gym}{g_{Y\!M}}

\def\De{\textrm{D8}}
\def\DeB{\overline{\textrm{D8}}}
\def\DeDeB{\textrm{D8-}\overline{ \textrm{D8}}}

\newcommand{\Ukk}{U_{\rm KK}}
\newcommand{\Mkk}{M_{\rm KK}}
\newcommand{\wt}{\widetilde}
\newcommand{\wh}{\widehat}

\newcommand{\Az}{\cala_{0}}
\newcommand{\dAz}{\dot{\cala_{0}}}
\newcommand{\dZAz}{\partial_Z \cala_{0}}

\newcommand{\dmuup}{{\partial^\mu}}

\newcommand{\bvo}{{{v}_0}}

\begin{document}

\begin{titlepage}


\begin{center}
  {\large \bf The Chiral Model of Sakai-Sugimoto at Finite Baryon Density} \\
\vspace{10mm}
  Keun-Young Kim$^a$, Sang-Jin Sin $^{a,b}$  and Ismail Zahed$^a$\\
\vspace{5mm}
$^a$ {\it Department of Physics and Astronomy, SUNY Stony-Brook, NY 11794}\\
$^b$ {\it Department of Physics, BK21 division, Hanyang University, Seoul 133-791, Korea}\\
\vspace{10mm}
\end{center}

\begin{abstract}
In the context of holographic QCD we analyze Sakai-Sugimoto's
chiral model at finite baryon density and zero temperature. The
baryon number density is introduced through compact D4 wrapping
$S^4$ at the tip of D8-$\DeB$. Each baryon acts as a chiral
point-like source distributed uniformly over $\mathbb{R}^3$, and
leads a non-vanishing $U(1)_V$ potential on the brane. For fixed
baryon charge density $n_B$ we analyze the   energy density and
pressure using the canonical formalism. The baryonic matter with
point like sources is always in the spontaneously broken phase of
chiral symmetry, whatever the density. The point-like nature of
the sources and large $N_c$ cause the matter to be repulsive as
all baryon interactions are omega mediated. Through the induced
DBI action on D8-$\DeB$, we study the effects of the fixed baryon
charge density $n_B$ on the pion and vector meson masses and
couplings. Issues related to vector dominance in matter in the
context of holographic QCD are also discussed.
\end{abstract}

\end{titlepage}

\renewcommand{\thefootnote}{\arabic{footnote}}
\setcounter{footnote}{0}

\newpage


\section{Introduction}

Dense hadronic matter is of interest to a number of fundamental
problems that range from nuclear physics to astrophysics. QCD at
finite baryon density is notoriously difficult: (1) the
introduction of a chemical potential causes most lattice
simulations to be numerically noisy owing to the sign problem; (2)
the baryon-baryon interaction is strong making most effective
approaches limited to subnuclear matter densities.

In the limit of a large number of colors $N_c$, QCD is an effective
theory of solely mesons where baryons appear as chiral skyrmions.
Dense matter in large $N_c$ is a skyrmion crystal with spontaneous
breaking of chiral symmetry at low density, and restored or stripped
(Overhauser) chiral symmetry at high density. While some of these
aspects can be studied qualitatively using large $N_c$ motivated
chiral models~\cite{DENSESKYRMION}, they still lack a first principle
understanding.

The AdS/CFT approach has provided a framework for discussing
large $N_c$ gauge theories at strong coupling $\lambda=g^2N_c$
from first principles~\cite{Maldacena}. A particularly interesting
AdS/CFT construction is the holographic and chiral approach
proposed by Sakai and Sugimoto~\cite{Sakai1, Sakai2} (SS model).
In the limit where $N_f\ll N_c$, chiral QCD is obtained as a
gravity dual to $N_f$ D8-$\DeB$ embedded into a $D4$ background
in 10 dimensions where supersymmetry is broken by the
Kaluza-Klein (KK) mechanism. The KK scale plays the role of the
chiral scale. The SS model yields a first principle effective
theory of pions, vectors, axials and baryons that is in good
agreement with experiment~\cite{Sakai1,Sakai2,Sakai3}. The SS
model at finite temperature has been studied in~\cite{FiniteT}
and the bayrons in the context of Skyrmion and Instanton have
been worked out~\cite{Sakai3, Baryons}.

Recently we have suggested that the SS model can be used to
analyze dense hadronic matter at large $N_c$ and strong coupling
$\lambda$ \cite{KSZ}. At zero temperature, the quark/baryon
chemical potential $\mu$ (or $\mu_B=m_B+N_c\mu$)
is introduced as the boundary value of
the $U(1)_V$ brane potential $A_0$.
The diagonalization of the vector modes in dense matter, enforces
vector dominance and  yields ${\cal A}_0=\mu\psi_v$ throughout
where $\psi_v$ is given in~\cite{Sakai2}. While the mode
decomposition of $\psi_v$ on the brane leads to a highly oscillating
${\cal A}_0$, we have argued in \cite{KSZ} that only those modes
in ${\cal A}_0$ below the KK scale should be retained. As a result
both brane  and meson properties in holographic and dense baryonic
matter were discussed.

Soon after the posting of this work, several studies appeared
addressing the same issue of baryonic matter in holographic QCD
including also temperature. In~\cite{Horigome} it was suggested
that the ${\cal A}_0$ field is instead fixed by the   equation
of motion on the brane by varying the pertinent DBI action. In the absence of
 brane ``charges'' the authors in~\cite{Horigome} concluded that
only a constant ${\cal A}_0$ is a solution, with no baryonic
effect at zero temperature. However, the conserved baryonic
charge has to be mirrored by the ``charge''  of baryon
vertex. The latter is obtained by considering
brane wrapping of $S^4$ within D8-$\DeB$. The wrapping number is
the bulk conserved''charge'' which is at the origin of a
non-constant ${\cal A}_0$ in bulk.

This point was further developed in~\cite{Sin1} and used to discuss
the phase structure of dense and hot holographic matter, albeit
in a brane set up without chiral symmetry. In~\cite{KMMMT} it
was argued that the additional ``charges'' in bulk upset the
smoothness of the DBI surface in bulk, leading to a spiky
structure due to the force balancing condition~\footnote{
Notice that  while the baryonic charges are space separated,
they occupy the same position in the dual transverse space!}.
As a result, the embedded branes should {\it always} touch the
horizon even for arbitrarily small temperature and/or density,
thereby altering totally the phase diagram in~\cite{Sin1}. This
latter point is physically unintuitive. Indeed, the ``charge''
or baryon vertex exists even at zero temperature with no need
to connect to any horizon. At finite temperature through the
insertion of a black hole this point is developed in details
in~\cite{Sin2}.

In this paper we follow on the analysis in~\cite{Sin1} in bulk
and at zero temperature (and low temperature before deconfinement phase transition).
The dual ``charge'' or baryon vertex is inserted at the tip of the minimally embedded D8-$\DeB$ surface.
This way each of D8, $\DeB$ is shared equally, leading to a {\it
chiral} baryon vertex.
Since no connecting string is involved,  there is no spiky structure involved here.
Also, there is a one-to-one correspondence between the baryon vertex
normalization and the Wess-Zumino term in the induced chiral DBI
action. Thus the boundary chiral skyrmions constructed from the
induced DBI action, are dual to {\it static} and {\it point-like}
instantons in bulk.
These conditions will be relaxed in a sequel. In many ways, this
approach complements the original discussion in~\cite{KSZ}.

In section 2, we briefly review the SS model and set up the
notations. In section 3, we introduce the $U(1)_V$ field ${\cal
A}_0$ in bulk and show how the baryon charge density $n_B$ affects
its minimal profile. In section 3 and 4, we construct the bulk
hamiltonian and derive the energy density as a function of the
identified baryon density. The energy density is found to grow
about quadratically with the baryon density. In section 5, we
summarize the construction of the chiral effective action for
pions, vectors and axials at zero density. In section 6, we show
how this chiral effective action is modified by the finite
``charges'' in bulk. A number of meson properties are discussed
as a function of the identified baryon number. Our conclusions
are in section 7. Throughout, the canonical formalism will be
used.

\section{SS Model}
In this section we summarize the D4/D8-$\DeB$ set up for notation
and completeness. For a thorough presentation we refer to~\cite{Sakai1}
and references therein.
The metric, dilaton $\phi$, and the 3-form RR field $C_3$ in $N_c$
D4-branes background are given by
\begin{eqnarray}
&&ds^2=\left(\frac{U}{R}\right)^{3/2} \left(\eta_{\mu\nu}dx^\mu
dx^\nu+f(U)d\tau^2\right) +\left(\frac{R}{U}\right)^{3/2}
\left(\frac{dU^2}{f(U)}+U^2 d\Omega_4^2\right)\  ,
\nn\\
&&~~~~e^\phi= g_s \left(\frac{U}{R}\right)^{3/4}, ~~F_4\equiv
dC_3=\frac{2\pi N_c}{V_4}\epsilon_4 \ , ~~~f(U)\equiv
1-\frac{\Ukk^3}{U^3} \ , \label{D4sol}
\end{eqnarray}
where $x^\mu= x^{0,1,2,3}$, $\tau (\equiv x^4)$ is the compact
variable on $S^1$. $U (\geq U_{KK})$ and $\Omega_4$ are the radial
coordinate and four angle variables in the $x^{5,6,7,8,9}$
direction. $R^3 \equiv \pi g_s N_c l_s^3$, where $g_s$ and $l_s$
are the string coupling and length respectively. $V_4=8\pi^2/3$
is the volume of unit $S^4$ and $\epsilon_4$ is the corresponding
volume form.

To avoid a conical singularity at $U=\Ukk$ the period of $\d\t$ of
the compactified $\t$ direction is set to
\begin{eqnarray}
\d\t = \frac{4\pi}{3}\frac{R^{3/2}}{\Ukk^{1/2}} \ .
\end{eqnarray}
in terms of which we define the Kaluza-Klein mass as
\begin{eqnarray}
\Mkk \equiv
 \frac{2\pi}{\d\t} = \frac{3}{2}
\frac{\Ukk^{1/2}}{R^{3/2}} \ .
\end{eqnarray}
The parameters $R$, $\Ukk$, and $g_s$ may be expressed in terms of
$\Mkk$, $\l( = \gym N_c)$, and $l_s$ as
\begin{eqnarray}
R^3 = \half \frac{\l  l_s^2}{\Mkk}\ , \quad \Ukk = \frac{2}{9} \l
\Mkk l_s^2 \ , \quad g_s = \frac{1}{2\pi}\frac{\l}{\Mkk N_c l_s} \
\end{eqnarray}
Now, consider $N_f$ probe D8-branes in the $N_c$ D4-branes
background. With $U(N_f)$ gauge field $A_M$ on the D8-branes, the
effective action consists of the DBI action and the Chern-Simons
action:
\begin{eqnarray}
S_{\De}&=& S^{DBI}_{\De} + S^{CS}_{\De}\ , \label{Action.0} \nn \\
S^{DBI}_{\De}&=& -T_8 \int d^9 x \ e^{-\phi}\ \Tr
\sqrt{-\det(g_{MN}+2\pi\alpha' F_{MN})} \ , \label{DBI}\\
S^{CS}_{\De}&=&\frac{1}{48\pi^3} \int_{D8} C_3 \Tr F^3 \label{CS}
\ .
\end{eqnarray}
where $T_8 = 1/ ((2\pi)^8 l_s^9)$, the tension of the D8-brane,
$F_{MN}=\partial_M A_N -\partial_N A_M -i \left[ A_M , A_N
\right]$ ($M,N = 0,1,\cdots,8$), and $g_{MN}$ is the induced
metric on  D8-branes:
\begin{eqnarray}
ds^2=\left(\frac{U}{R}\right)^{3/2}\!\!\!\!  \eta_{\mu\nu}dx^\mu
dx^\nu + \left[ \left(\frac{U}{R}\right)^{3/2}\!\!\!\! f(U)
(\tau'(U))^2 +\left(\frac{R}{U}\right)^{3/2}\!\!\!\!
\frac{1}{f(U)} \right]dU^2+
\left(\frac{R}{U}\right)^{3/2}\!\!\!\! U^2 d\Omega_4^2 \ ,
\end{eqnarray}
where $\t' = \frac{d \t }{d U} $. The effective action of $\DeB$
has the same form and the total action of $N_f$ D8-$\DeB$-branes
has a symmetry
\begin{eqnarray}
U(N_f)_L \times U(N_f)_R = SU(N_f)_L \times SU(N_f)_R \times
U(1)_V \times U(1)_A \ ,
\end{eqnarray}
which is interpreted as a flavor chiral symmetry of massless
quarks.

In the SS model baryons are skyrmions in $\mathbb{R}^3$. However,
in the gravity dual they are either an effective fermion degree
of freedom in 5-dimensions (bottom-up) or an instanton wrapping
$D4$ (top-down) and sourcing the baryon current through the
Chern-Simons term \cite{ Sakai3, KSZ, Harvey, HRYY}. In both
cases, the baryon can be treated as a delta function source of
the  brane gauge field ${\cal A}_0$ (which is $\omega_0$ in
$\mathbb{R}^3$), which we will use explicitly.

\section{Background field $\Az$} \label{Sec:Chemical}
 Let $\Az(U)$ be a $U(1)_V$ valued background gauge
field in bulk. Its boundary value is related to the baryon
chemical potential \cite{KSZ, Horigome, Sin1, Sin2}. In the
absence of the source,  the effective action of the D8-branes (\ref{Action.0})
becomes
\begin{equation}
S_{\De} =
 -\frac{N_f T_8 V_4}{g_s} \int d^4x 
\, dU U^4 \left[ f \, (\tau')^2 + \left( \frac{R}{U} \right)^3
\left( f^{-1} - \left( 2\pi\alpha' \Az' \right)^2 \right)
\right]^{\frac{1}{2}}, \label{Action.1}
\end{equation}
where $\Az' = \frac{d\Az}{dU}$ and the Chern-Simons action
vanishes. The equations of motion for $\tau(U)$ and $\Az(U)$ are
\cite{Horigome}
\begin{eqnarray}
\frac{d}{dU} \left[ \frac{U^4 f \, \tau'} {\sqrt{f \, (\tau')^2 +
\left( \frac{R}{U}\right)^3 \left( f^{-1} - \left( 2\pi\alpha'
\Az' \right)^2 \right) }} \right]  =  0,
\end{eqnarray}
\begin{eqnarray}
\frac{d}{dU} \left[ \frac{U^4 \left( \frac{R}{U} \right)^3 \Az'}
{ \sqrt{ f \left( \tau' \right)^2 + \left( \frac{R}{U} \right)^3
\left( f^{-1} - \left( 2\pi\alpha' \Az' \right)^2 \right)} }
\right]  =  0.
\end{eqnarray}
In this paper we consider  only the case $\t' = 0$,  Sakai-Sugimoto's original embedding \cite{Sakai1, Sakai2},
where the D8-branes configuration in the $\t$ coordinate is not affected by
the existence of background $\Az$. This corresponds to $\t =
\frac{\d\t}{4}$, the maximal asymptotic separation between D8 and
$\DeB$ branes.

To compare with \cite{Sakai1, Sakai2} we change the variable $U$
to $z$ through
\begin{eqnarray} \label{Uz}
U \equiv (\Ukk^3 + \Ukk z^2)^{1/3}
\end{eqnarray}
The action (\ref{Action.1}) is then
\begin{eqnarray}
S_{\De} =  -N_f \wt T \int d^4x \dint{0}{\infty}dz\, U^2  \, \sqrt
{\, 1 - (2\pi\alpha')^2 \frac{9}{4}\frac{U_z}{\Ukk} (\dell_z
\Az)^2 } \ , \label{Action.2}
\end{eqnarray}
where we used $\t' = 0$ and $\wt T \equiv \frac{N_c \Mkk}{216\pi^5
\a'^3}$. It is useful to define the dimensionless quantities
\begin{eqnarray}
Z \equiv \frac{z}{\Ukk}, \quad K(U) \equiv 1+ Z^2 =
\left(\frac{U}{\Ukk}\right)^3 \ ,
\end{eqnarray}
in terms of which the action is written as\footnote{The integral
is extended to $(-\infty, \infty)$ to take into account $\DeB$
branes as well as D8 branes.}
\begin{eqnarray}
S_{\De} =  -a \int d^4x \int dZ\, K^{2/3}  \, \sqrt {\, 1 - b
K^{1/3}
  (\dZAz)^2 } \ ,\label{Action.without.vector}
\end{eqnarray}
where
\begin{eqnarray}
a \equiv \frac{N_c N_f \l^3 \Mkk^4}{3^9 \pi^5} \ , \quad \quad b
\equiv \frac{3^6 \pi^2}{4 \l^2 \Mkk^2} \ .
\end{eqnarray}

Now we introduce the baryon source coupled to $\cala_0$ through
the Chern-Simons term \cite{KSZ, Sakai3, Harvey}  as mentioned
before. We assume that baryons are uniformly distributed over
$\mathbb{R}^3$ space whose volume is $V$. For large $\lambda$,
the instanton size is $1/\sqrt{\lambda}$ \cite{Sakai3,HRYY}. It
can be treated as a static delta function source at large $N_c$.
For a uniform baryon distribution, the source is
 \be
   S_{\mathrm{source}}= N_c n_B \int d^4x \int dZ\, \delta (Z)\cala_0(Z).\ee
  The equation of motion of $\cala_0$ is
\be \frac{d }{dZ} \frac{\partial {\cal L}}{\partial (\partial_Z
\cala_0)}
 = n_q \delta(Z)\ , \label{Source.action}
\ee
which yields
\begin{eqnarray}
\frac{\partial {\cal L}}{\partial (\partial_Z \cala_0)}
 = \half n_q \ \mathrm{sgn}(Z)\ , \label{EqnA0}
\end{eqnarray}
where  $n_q = N_c n_B$  is  the quark density and  the step
function  $ \mathrm{sig}(Z)$ is determined by the  symmetry
between  D8 ($Z>0$) and $\DeB (Z<0)$. By integrating once more we
get the classical solution $\Az$
\begin{eqnarray}
\Az(Z;n_q) = \Az(0)+ \dint{0}{Z}dZ \frac{n_q/2}{\sqrt{{(ab)^2K^2}
+ b K^{1/3}n_q^2/4 }}\ .\label{Az.solution}
\end{eqnarray}
\begin{figure}[]
  \begin{center}
   \subfigure[] {\includegraphics[width=8cm]{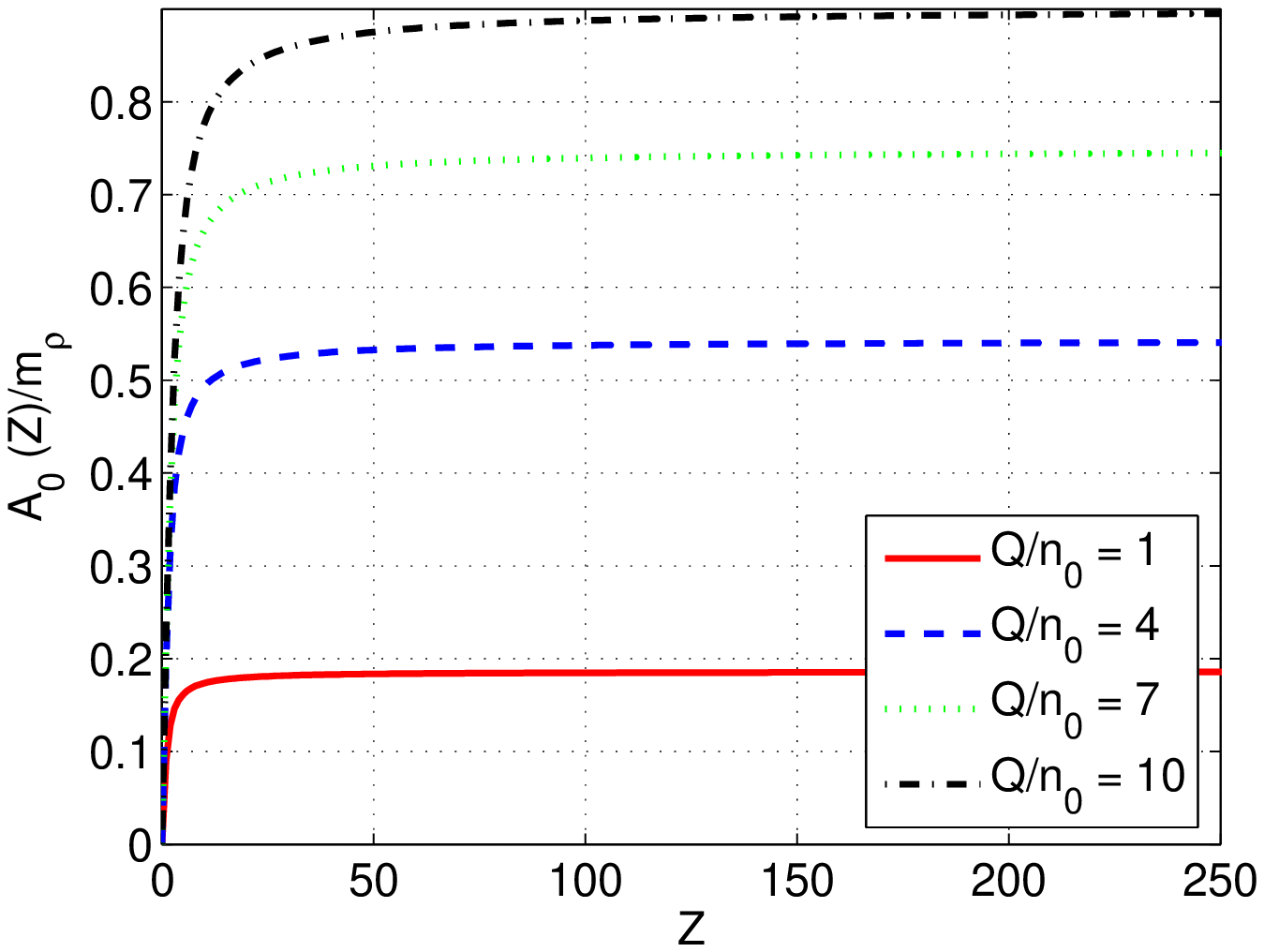}}
   \subfigure[] {\includegraphics[width=8cm]{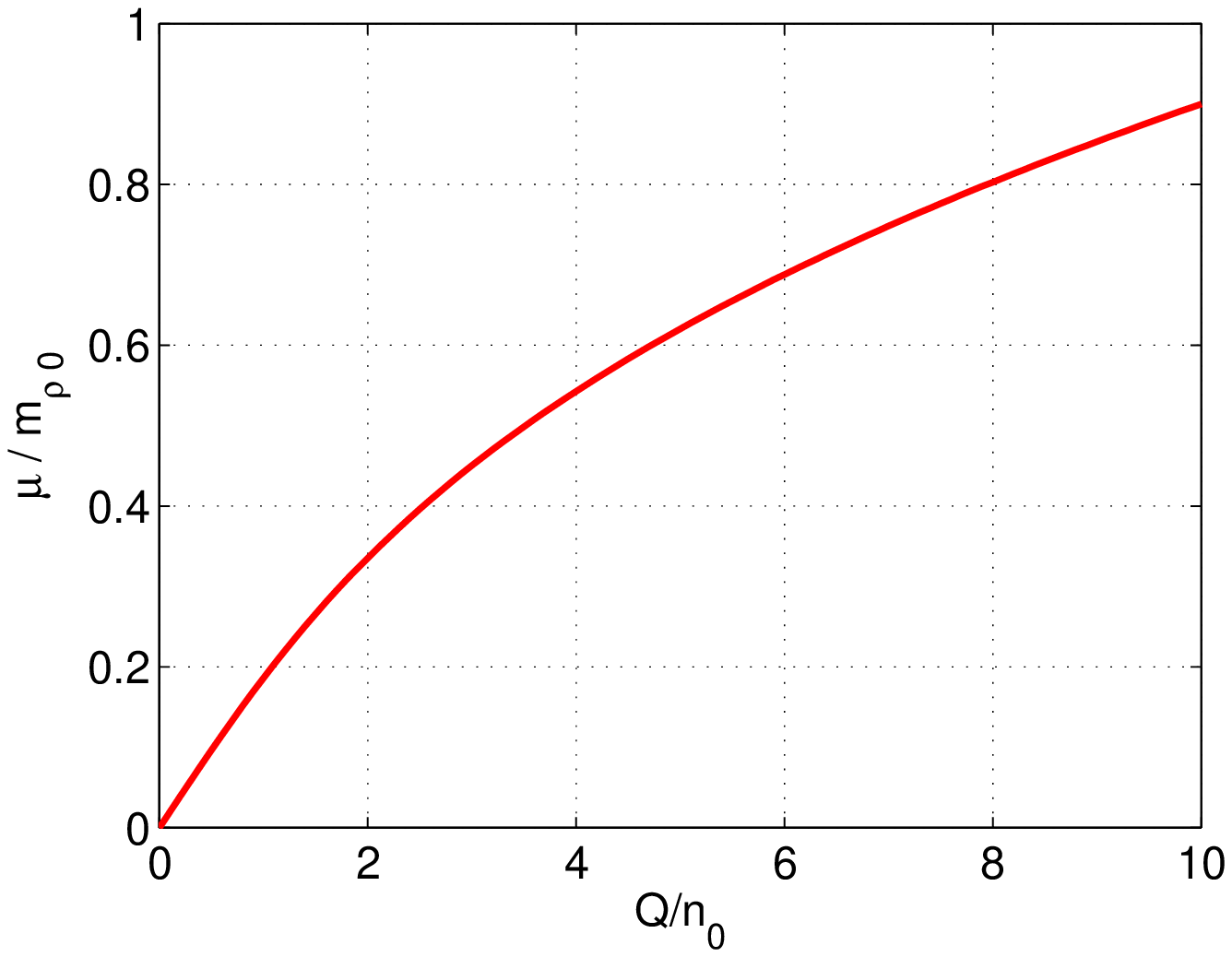}}
  \caption{(a) The profile of $\Az(Z)$, (b) Chemical potential vs baryon charge ($ \frac{\mu}{m_\rho}$
vs $\frac{Q}{n_0}$, where $Q \equiv n_q/2$).}
  \label{Fig:A0}
  \end{center}
\end{figure}

We introduce the  ``baryon charge chemical potential of a
quark'', $\m$,  by ~\cite{Sin1,Sin2}
\begin{eqnarray}
\mu(n_q) \equiv \lim_{|Z|\ra \infty} \Az(Z;n_q)  \ . \label{Mu.Q}
\end{eqnarray}
This relation also defines $\m$ as a function of $n_q$ and vice
versa. Furthermore we define the baryon chemical potential as
\begin{eqnarray}
\m_B = m_B + N_c \m \ .
\end{eqnarray}
In Fig.(\ref{Fig:A0}a) we plot the profile of $\Az(Z)$ in the $Z$
coordinate and in Fig.(\ref{Fig:A0}b) we show $\m$ for various
baryon densities. Since we work in the canonical formalism $\m$
is more like a Lagrange constraint.

Throughout, the numerics will be carried using the
following values~\cite{Sakai1,Sakai2}:
$N_f=2$, $N_c = 3$, $f_\pi = 92.6$MeV, and $m_\rho =
776$MeV. The smallest eigenvalue was calculated to
be $\l_1 = 0.669$. Using these five values we can estimate
$\Mkk$, $\l$, $\kappa$, $a$, and $b$:
 \be
 \Mkk = \frac{m_\rho}{\sqrt{\l_1}} \simeq 950 MeV,  \;\; \l
\equiv \gym^2 N_c = f_\pi^2\frac{54\pi^4}{N_c \Mkk^2} \simeq
16.71 ,\;\; \kappa \equiv  \frac{\l N_c}{216 \pi^3} \simeq
0.0075, \ee and \be \quad a=3.76\cdot10^9 MeV^4, \quad
b=7.16\times 10^{-6}MeV^{-2}.\ee The definition of $\k$ and $\l$
are different from \cite{Sakai1, Sakai2} by a factor of $2$, but
it is consistent with \cite{Sakai3}. In all figures $n_B$ is
normalized to $\frac{n_B}{n_0}$, with $n_0$ the nuclear matter
density,
\be
n_0 = 0.17 fm^{-3} \simeq 1.3\times 10^6 MeV^3.\ee

\section{Thermodynamics} \label{Sec:Thermodynamics}
Consider the action (\ref{Action.without.vector}) with the source
term (\ref{Source.action}),
\begin{eqnarray}
&&S = \int d^4x \dint{-\infty}{+\infty}dZ\, \call \nn \\
&&\quad \textrm{ with }\quad \call \equiv - a  K^{2/3}  \, \sqrt
{\, 1 - b K^{1/3}
  (\dZAz)^2 }  + n_q \d(Z)\cala_0(Z) \ .\label{E.Action.without.vector}
\end{eqnarray}
The $\Az$ is an auxillary field with no time-dependence. It can
be eliminated by the equation of motion (\ref{EqnA0}) and
(\ref{Az.solution}). The energy is
\begin{eqnarray}
U (n_q) &=& \int dx^3\int_{-\infty}^{+\infty} dZ\, ( - \call) \nn\\
&=&  a V \int_{-\infty}^{+\infty} dZ\,  K^{2/3}\sqrt{1 +
\frac{n_q^2}{ 4 a^2 b}K^{-5/3}}  -n_q \mu \ ,
\end{eqnarray}
where $V$ is short for $\int dx^3$ and we may set $\mu=0$. The
chemical potential $\mu$ is constrained by the Gibbs relation
$\mu = \frac{\dell F(n_q)}{\dell n_q}$ where $F(n_q)$ is the
Helmholtz free energy  which is $U(n_q)$ at zero temperature. Thus
\begin{eqnarray}
&& \m = \dint{-\infty}{\infty}dZ \frac{n_q/4}{\sqrt{{(ab)^2K^2} +
b K^{1/3}n_q^2/4 }}\  \label{Mu.check} \ ,
\end{eqnarray}
which is in agreement with the solution (\ref{Az.solution}) for
$A_0(0)=0$. We note that this construction is consistent with
\cite{KSZ,Horigome,KMMMT,Sin1} where the grand potential is
identified with the DBI action at finite $\m$.

In terms of the baryon number density $n_B$ (${n_q}/{N_c}$) the
regularized Helmholtz free energy is
\begin{eqnarray}
\frac{F_{\textrm{reg}}(n_B)}{a V}&\equiv&  \dint{-\infty}{\infty}
dZ K^{2/3} \left[  \sqrt {{1 + \frac{(N_c n_B)^2}{4 a^2
b}K^{-5/3}}}-1\right]  \ , \label{Reg.F}
\end{eqnarray}
after subtracting the vacuum value. The regularized internal
energy $U$, pressure $p$ and grand potential $\Omega$ as a
function of baryon number density $n_B$ or the baryon chemical
potential $\mu_B$ are
\begin{eqnarray}
\frac{U_{\textrm{reg}}(n_B)}{a V}&=&  \dint{-\infty}{\infty} dZ
K^{2/3} \left[  \sqrt{{1 + \frac{(N_c n_B)^2}{4 a^2 b}K^{-5/3}}}
-1\right] \
,\nn \\
\frac{p(n_B)_{\textrm{reg}}}{a} &=&   \dint{-\infty}{\infty} dZ\,
K^{2/3} \left[1 - \frac{1}{\sqrt {\, 1 + \frac{(N_c n_B)^2}{4 a^2 b} K^{-5/3} }} \right]\ , \nn \\
\frac{\Omega_{\textrm{reg}} (\wt{\m_B})}{a V} &=&
\dint{-\infty}{\infty} dZ\,
K^{2/3} \left[ \frac{1}{\sqrt {\, 1 + \frac{(N_c n_B(\wt{\m_B}))^2}{a^2 b} K^{-5/3} }} - 1 \right]\ , \nn \\
\wt{\m_B} &=& N_c \dint{-\infty}{\infty}dZ \frac{N_c
n_B/4}{\sqrt{{(ab)^2K^2} + b K^{1/3}(N_c n_B/2)^2 }} \ ,
 \label{Plot.functions}
\end{eqnarray}
where $\wt{\m_B} \equiv \m_B -  m_B = N_c \m$.
\begin{figure}[]
  \begin{center}
   \includegraphics[width=8cm]{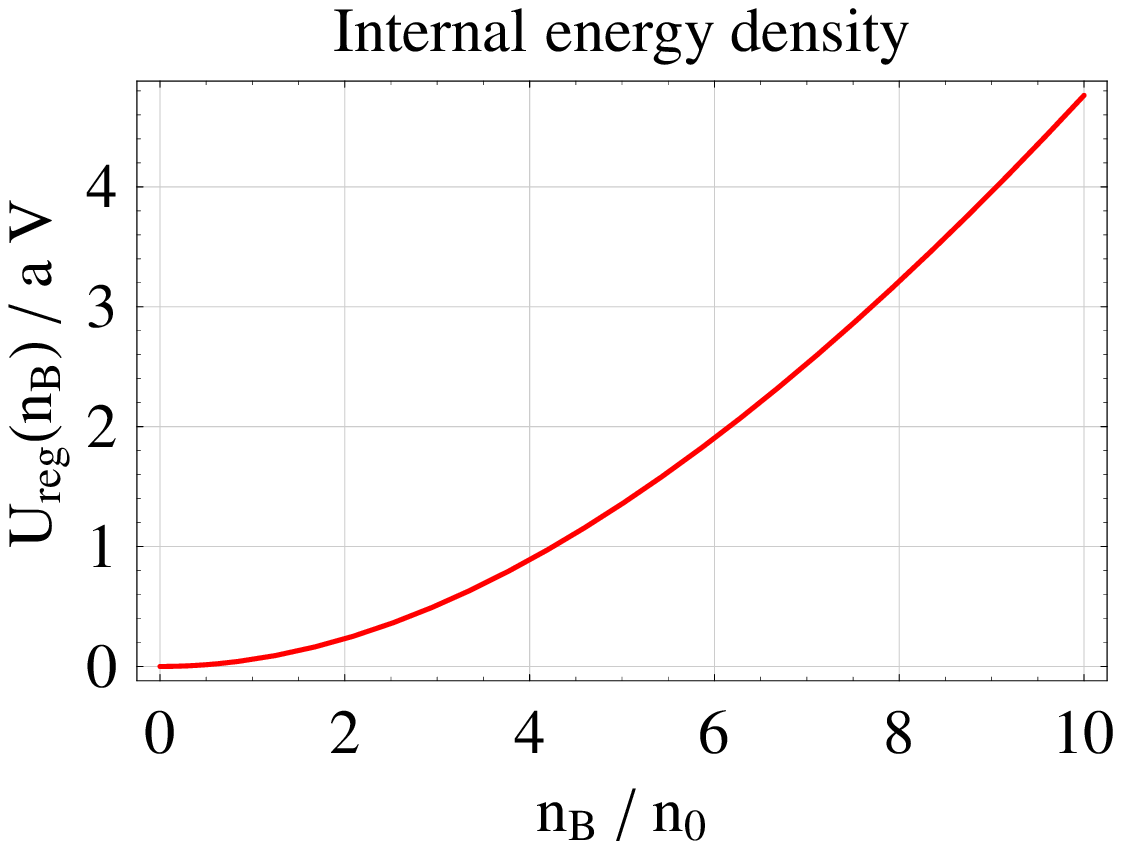}
   \includegraphics[width=8cm]{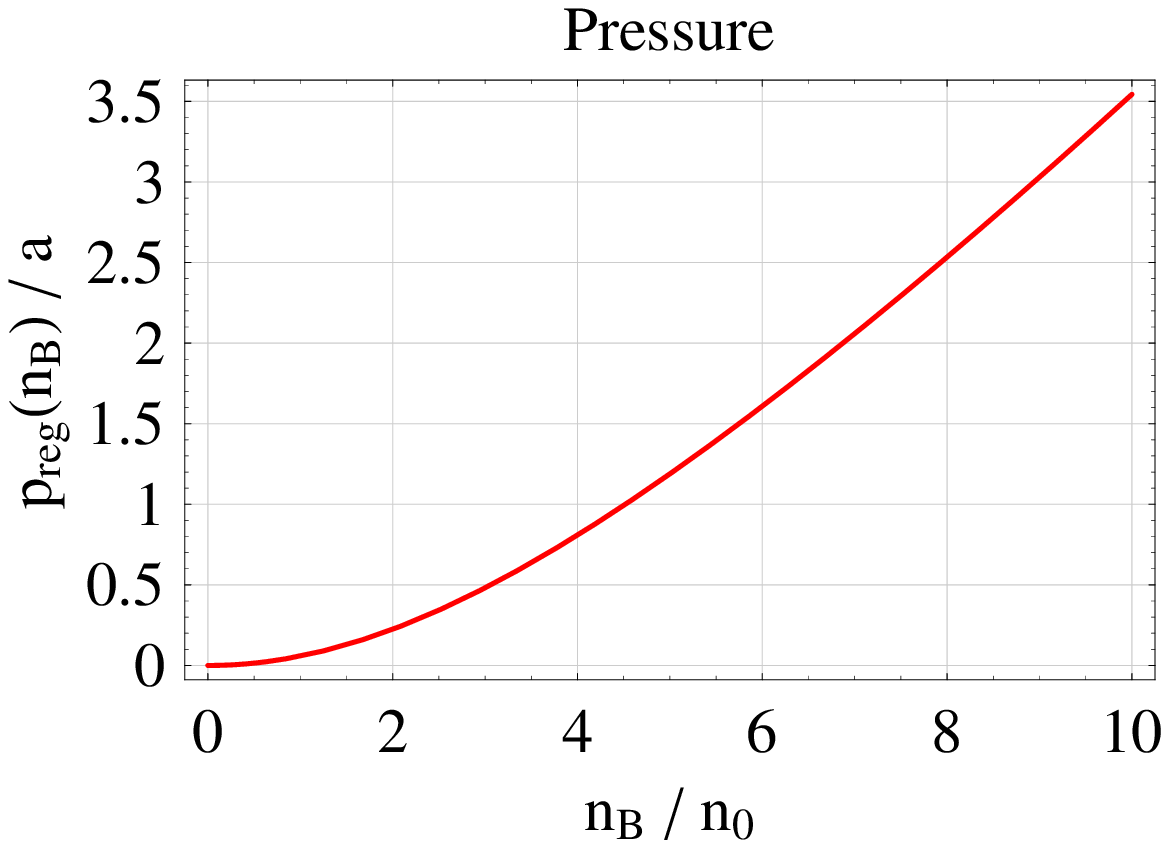}
   \includegraphics[width=8cm]{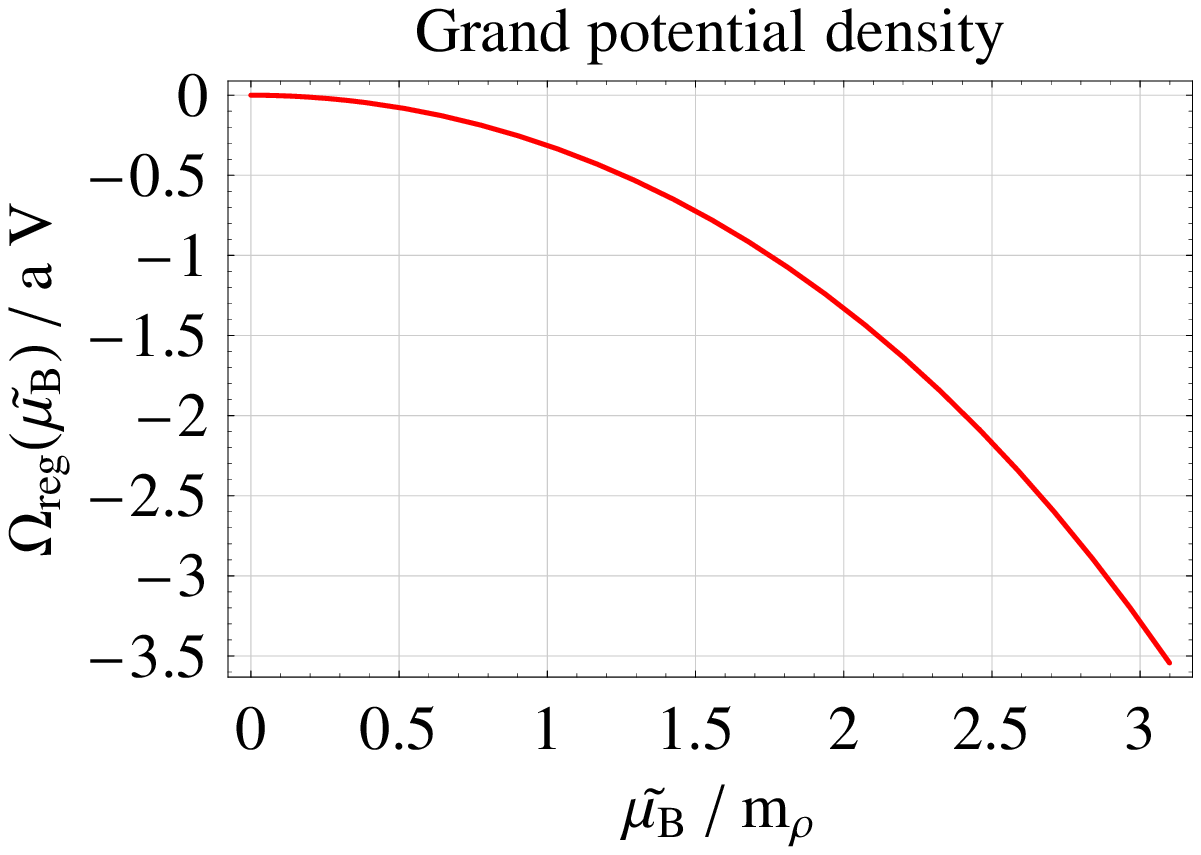}
   \includegraphics[width=8cm]{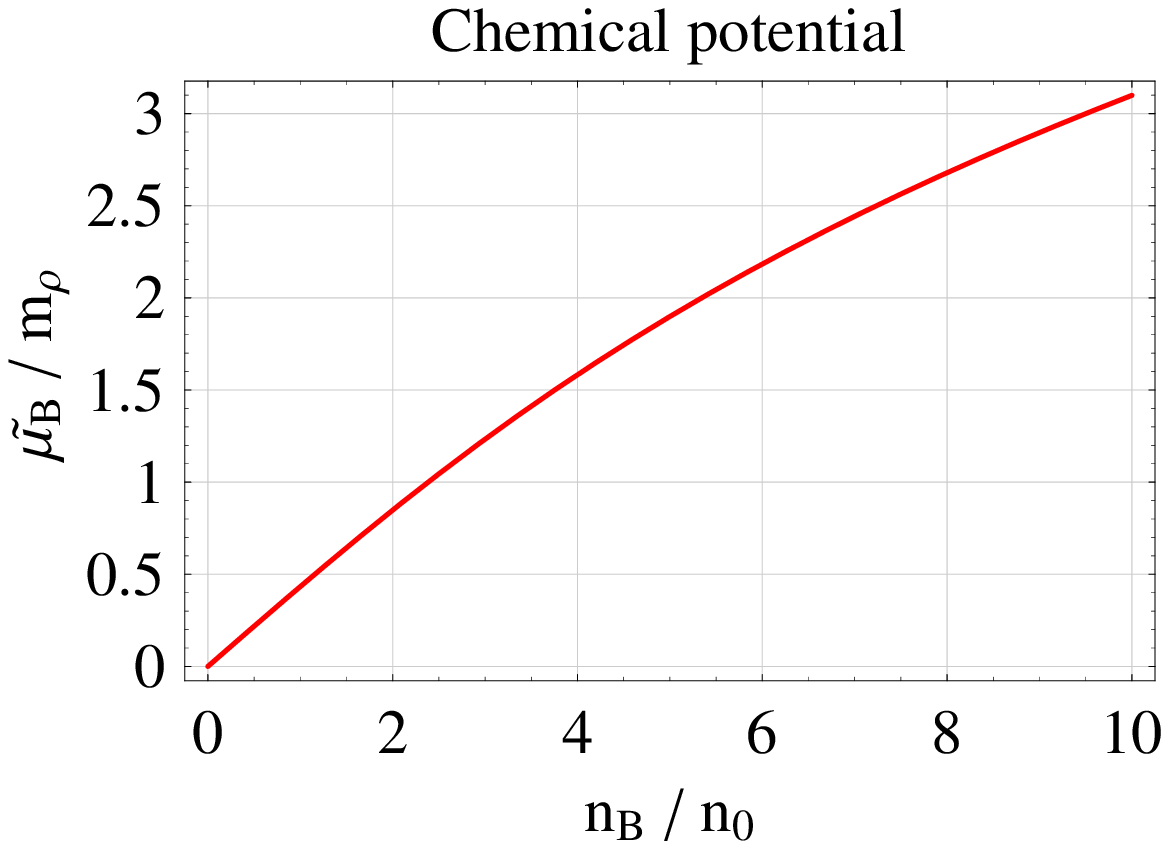}
  \caption{Numerical behaviour of the thermodynamic functions: See Eq.(\ref{Plot.functions})}
  \label{Fig:Thermodynamic.potential}
  \end{center}
\end{figure}

In Fig.(\ref{Fig:Thermodynamic.potential}) we present the
numerical plots of these thermodynamic functions with the
numerical inputs in section \ref{Sec:Chemical}. For small baryon
densities the energy density is quadratic in $n_B/n_0$ (or
$\mu/m_\rho$). At large baryon densities it is of order
$(n_B/n_0)^{1.4}$. The small density limit can be qualitatively
understood by noting that in bulk the ${\cal A}_0$ configuration
for fixed charge is obtained by minimizing the induced DBI action of
D8-$\DeB$. Thus only flavor-meson mediated interactions between
the point-like baryons are included.
\begin{table}[!b]
\begin{center}
\begin{tabular}{|c|c|c|c|}
\hline
\vspace{-0.4cm}& & &  \\
Thermodynamic function &  $n_B/n_0 \sim 0$ & $n_B/n_0 \sim 10$ &$n_B/n_0 \ra \infty   $\\
\vspace{-0.4cm}& & &  \\
\hline \hline
\vspace{-0.4cm}& & &  \\
Internal energy & $ (n_B/n_0)^2 $ & $ (n_B/n_0)^{1.85} $ &  $(n_B/n_0)^{1.4}$  \\
\vspace{-0.4cm}& & &  \\
Pressure & $ (n_B/n_0)^2 $ & $ (n_B/n_0)^{1.45} $& $ (n_B/n_0)^{1.4} $   \\
\vspace{-0.4cm}& & &  \\
Chemical potential & $ (n_B/n_0)^1$ & $ (n_B/n_0)^{0.67}  $&  $ (n_B/n_0)^{0.4}  $   \\
\vspace{-0.4cm}& & &  \\
Grand potential& $ -(\wt{\m_B}/m_\rho)^{2}  $ & $
-(\wt{\m_B}/m_\rho)^{2.16}  $ & $ -(\wt{\m_B}/m_\rho)^{3.5}  $
\\ \hline
\end{tabular}
\end{center}
   \caption{Numerical behaviour of the thermodynamic functions:
    See Eq.(\ref{Plot.functions}) }
    \label{Thermosummary}
\end{table}
At large $N_c$ the $D4$ mediated correlated gravitons (glueballs
on the boundary) are heavy and decouple. Since our point baryonic
vertices in bulk map on infinite size skyrmions at the boundary
this implies that only $\omega$ exchanges survive at large $N_c$.
Rho and pion exchange relies on skyrmion gradients which are zero. At low
baryon densities, the dominant Skyrmion-omega-Skyrmion
interaction is two-body and repulsive. Thus the energy density is
positive and quadratic in the baryon density. The baryonic matter
is prevented from flying apart by the container $V$. At large
baryon densities, the energy density softens as the quark chemical
potential is seen to saturate to $(n_B/n_0)^{0.4}$ numerically.
We recall that the baryons are fixed sources so no Fermi motion
is involved to this order. The pressure behaves as $(n_B/n_0)^2$
at low baryon densities, and again softens to $(n_B/n_0)^{7/5}$
at large baryon densities from the plot.  We summarize the
behaviour of the thermodynamic functions obtained numerically in
Table.(\ref{Thermosummary}). In this paper we do not consider the
back reaction of gravity for baryons or D8 brane, therefore
 the behaviour at higher densities, say $n_B/n_0 \gg 10$, is not
justified.

\section{Effective meson action: $n_B = 0$}

In~\cite{Sakai1, Sakai2} the meson spectrum and coupling was
studied at zero baryon density by analyzing the DBI action of
D8-$\DeB$ branes with the fluctuating gauge field $A_M$. We want
to extend the analysis to {\it finite baryon density} or $n_B \neq
0$. For this purpose we streamline in this section the
construction in~\cite{Sakai1, Sakai2} for notational purposes and
completeness. In the next two sections we add the background
$U(1)_V$ field $\Az$ to the fluctuating gauge field $A_M$. It
will enable us to study meson properties at finite baryon density.

\subsection{Mode decomposition of $A_M$}
The gauge field $A_M$ has nine components, $A_\mu = A_{1,2,3,4}$, $A_z
(\equiv A_5)$, and $A_{\a}$($\a = 5,6,7,8$, the coordinates on the
$S^4$). We assume that $A_{\a} = 0$, and $A_\mu$ and $A_z$ are
independent of the coordinate on $S^4$. We further assume that
$A_M$ can be expanded in terms of complete sets, $\psi_n(z)$ and
$\phi_n(z)$ as
\begin{eqnarray}
A_\mu(x^\mu,z)&=& \sum_{n=1}^\infty B_\mu^{(n)}(x^\mu)\psi_n(z)\ ,
\label{Amu.1}\\
A_z(x^\mu,z)&=&\varphi^{(0)}(x^\mu)\phi_0(z)+ \sum_{n=1}^\infty
\varphi^{(n)}(x^\mu)\phi_n(z)\ , \label{Az.1}
\end{eqnarray}
where $B_\mu^{(n)}$ is identified with vector and axial vector
mesons and $\varphi^{(0)}$ with pions. $\varphi^{(n)}$ can be
absorbed into $B_\mu^{(n)}$ through the gauge transformation
(section \ref{sec.meson.action}). $\psi_n$ satisfies the
eigenvalue equation,
\begin{equation}
-K^{1/3}\,\dell_Z\left( K\,\dell_Z\psi_n\right)= \lambda_n\psi_n
\ , \label{psi}
\end{equation}
with the boundary condition $ \dell_Z \psi_n (0) = 0 $ (vector
meson) or $ \psi_n (0) = 0 $ (axial vector meson) at $Z = 0$.
They are normalized by
\begin{eqnarray}
\kappa \int dZ \,K^{-1/3}\psi_n \psi_m = \d_{nm} \ ,
\label{psi:norm}
\end{eqnarray}
where $ \kappa \equiv \wt T (2\pi \alpha')^2 R^3 =\frac{\lambda
N_c}{216\pi^3}\ $, and (\ref{psi}) and (\ref{psi:norm}) implies
\begin{eqnarray}
\kappa \int dZ \, K ( \dell_Z \psi_n)( \dell_Z \psi_m) =\lambda_n
\d_{nm} \ . \label{dpsi}
\end{eqnarray}
The $\phi_n(Z)$ are chosen such that
\begin{eqnarray}
\phi_n(Z)  &=& \frac{1}{\sqrt{\l_n}\Mkk U_{KK}} \dell_Z\psi_n(Z)\quad (n\ge 1) \ , \\
\phi_0(Z)  &=& \frac{1}{\sqrt{\pi\kappa}M_{kk}U_{KK}}\inv{K} \ ,
\label{phi.0}
\end{eqnarray}
with the normalization condition:
\begin{eqnarray}
(\phi_m,\phi_n)\equiv \k \Mkk^2 \Ukk^2 \int dZ \,K\,\phi_m \phi_n
=\delta_{mn} \ , \label{phi:ortho}
\end{eqnarray}
which is compatible with (\ref{dpsi}).

\subsection{Effective meson action}\label{sec.meson.action}
With the gauge field $A_\mu(x^\m,z)$ and $A_z(x^\m,z)$ the DBI
action of the D8-$\DeB$-branes becomes
5-dimensional~\footnote{The gauge group generators $t^a$ are
normalized as $\Tr t^a t^b = \d_{ab}/2$}:
\begin{eqnarray}
&& S_{\DeDeB}^{DBI}
= -\wt T \int d^4x dz\, U^2  \nn \\
&&~~~~~~~~ \Tr \, \sqrt {\, 1 + (2\pi\alpha')^2 \frac{R^3}{2U^3}
 F_{\mu \nu}F^{\mu \nu}
+ (2\pi\alpha')^2 \frac{9}{4}\frac{U}{\Ukk}  F_{\mu z}F^{\mu z} +
[F^3] + [F^4] + [F^5]  } \ , \label{DBI.1}
\end{eqnarray}
where $\wt T = \frac{N_c}{216\pi^5}\frac{M_{KK}}{\alpha'^3}$, $U$
is a function of $z$ by (\ref{Uz}), and the indices are
contracted by the metric $(-,+,+,+,+)$. $[F^3],[F^4]$, and $[F^5]$
are short for the terms of $F^3, F^4$, and $F^5$ respectively.
Notice that the range of $z$ is extended from $[0,\infty]$ to
$[-\infty,\infty]$ to account for both $D8$ and $\DeB$.

Inserting (\ref{Amu.1}) and (\ref{Az.1}) into (\ref{DBI.1}) and
using the orthonomality of $\psi_n$ and $\phi_n$
((\ref{psi:norm})$\sim$(\ref{phi:ortho})), we have~\cite{Sakai1,
Sakai2}
\begin{eqnarray}
S_{\DeDeB}^{DBI}&\sim& \int d^4 x\,\Tr\left[
(\dell_\mu\varphi^{(0)})^2+\sum_{n=1}^\infty\left( \half
(\dell_\mu B^{(n)}_\nu-\dell_\nu B^{(n)}_\mu)^2 +\lambda_nM_{KK}^2
(B_\mu^{(n)}-\lambda_n^{-1/2}\dell_\mu\varphi^{(n)})^2
\right)\right]\nn\\
&&~~~~~+(\mbox{interaction terms})\ . \label{SSaction}
\end{eqnarray}
Here $\varphi^{(0)}$ and $B_\mu^{(n)}$ are interpreted as a masseless
pion field and an infinite tower of vector (or axial) vector meson
fields with masses $m_n^2 (\equiv \l_n \Mkk^2)$. The lightest vector
meson $\rho$ is identified with $B_\mu^{(1)}$. $\varphi^{(n)}$ are
absorbed into $B_\mu^{(n)}$. In the expansion (\ref{Amu.1}) and
(\ref{Az.1}), we have implicitly assumed that the gauge fields are
zero asymptotically, i.e. $A_M(x^\mu,z)\ra 0$ as $z\ra\pm\infty$.
The residual gauge transformation  that does not break this
condition is obtained by a gauge function $g(x^\mu,z)$ that
asymptotes a constant $g(x^\mu,z)\ra g_\pm$ at $z\rm \pm\infty$.
$(g_+,g_-)$ are interpreted as elements of the chiral symmetry group
$U(N_f)_L\times U(N_f)_R$ in QCD with $N_f$ massless flavors.

\subsection{$A_z = 0$ gauge and pion effective action}\label{A_z=0.gauge}

In the previous subsection we worked in the gauge $A_M(x^\m, z)
\ra 0 $ as $z \ra \pm \infty$. However the $A_z=0$ gauge can be
achieved by applying the gauge transformation $A_M\ra g A_M
g^{-1}+ g\dell_M g^{-1}$ with the gauge function
\begin{eqnarray}
g^{-1}(x^\mu,z)= P \exp\left\{- \int_{0}^z dz'\, A_z(x^\mu,z')
\right\} \ . \label{ginv}
\end{eqnarray}
Then the asymptotic values of $A_\m(z \ra \infty)$ do not vanish
and change to
\begin{eqnarray}
A_\m(x^\m,z) \ra \xi_\pm(x^\m)\dell_\m \xi_\pm^{-1}(x^\m) \qquad
\textrm{as }\ z \ra \pm \infty \ , \label{B.value}
\end{eqnarray}
where $\xi_\pm(x^\mu)\equiv \lim_{z\ra\pm\infty}g(x^\mu,z)$. The
gauge fields can be expanded as
\begin{eqnarray}
A_\mu(x^\mu,z) &=& \xi_+(x^\m)\dell_\m
\xi_+^{-1}(x^\m)(x^\mu)\psi_+(z) +\xi_-(x^\m)\dell_\m
\xi_-^{-1}(x^\m)\psi_-(z) +\sum_{n=1}^\infty
B_\mu^{(n)}(x^\mu)\psi_n(z)\ , \nn  \\
A_z(x^\mu,z) &=& 0 \ ,
\end{eqnarray}
where $\psi_\pm$ is the non-normalizable zero mode of (\ref{psi})
with the appropriate boundary condition to yield (\ref{B.value}):
\begin{eqnarray}
&& \psi_\pm = \half \pm \wh{\psi}_0 \ , \nn\\
&& \wh{\psi}_0 = \inv{\pi}\arctan(Z)
\end{eqnarray}

There is a residual gauge symmetry which maintains $A_z=0$. It is
given by the $z$-independent gauge transformation $h(x^\m)$,
\begin{eqnarray}
A_M(x^\m ,z) \ra h(x^\m)A_M(x^\m , z)h^{-1}(x^\m) + h(x^\m)\dell_M
h^{-1}(x^\m) \ ,
\end{eqnarray}
which acts on the component fields  as
\begin{eqnarray}
\xi_\pm&\ra& h\,\xi_\pm\, g_\pm^{-1}\ ,\label{xih}\\
B_\mu^{(n)}&\ra& h\,B_\mu^{(n)}\, h^{-1}\ , \label{tr2}
\end{eqnarray}
where we considered chiral symmetry $g_\pm$ together. Then
$\xi_\pm(x^\mu)$ are interpreted as the $U(N_f)$ valued fields
$\xi_{L,R}(x^\mu)$ which carry the pion degrees of freedom in the
hidden local symmetry approach . Indeed the transformation
property (\ref{xih}) is the same as that for $\xi_{L,R}(x^\mu)$
if we interpret $h(x^\mu)\in U(N_f)$ as the hidden local
symmetry. They are related to the $U(N_f)$ valued pion field
$U(x^\mu)$ in the chiral Lagrangian by
\begin{eqnarray}
\xi_+^{-1}(x^\mu)\xi_-(x^\mu)=U(x^\mu)\equiv
e^{2i\Pi(x^\mu)/f_\pi} \ .
\end{eqnarray}
The pion field $\Pi(x^\mu)$ is identical to $\varphi^{(0)}(x^\mu)$
in (\ref{Az.1}) in leading order.
A convenient gauge choice is
\begin{eqnarray}
\xi_-(x^\mu)=1, \quad
\xi_+^{-1}(x^\mu)=U(x^\mu)=e^{2i\Pi(x^\mu)/f_\pi}
\label{Hidden.gauge.1}
\end{eqnarray}
which expresses the gauge fields as,
\begin{eqnarray}
A_\mu(x^\mu,z) = U^{-1}(x^\mu)\dell_\mu U(x^\mu)\psi_+(z)
+\sum_{n\ge 1} B_\mu^{(n)}(x^\mu) \psi_n(z)
\label{Gauge.field.for.Skyrme}
\end{eqnarray}
In this gauge, after omitting the vector meson fields
$B_\m^{(n)}$, the effective action reduces to the Skyrme model
\begin{eqnarray}
S_{\DeDeB}^{DBI}\Big|_{B_\m^{(n)} =0} =\int d^4 x\left(\frac{\k
\Mkk^2 }{\pi} \Tr \left(U^{-1}\dell_\mu U\right)^2+
\frac{1}{32e_S^2}\Tr\left[U^{-1}\dell_\mu U,U^{-1}\dell_\nu
U\right]^2 \right) \ , \label{Skyrme}
\end{eqnarray}
where $e_S^{-2}\equiv\kappa \int\! dz\, K^{-1/3}(1-\psi_0^2)^2 $
and the pion decay constant $f_\pi$ is fixed by the comparison
with the Skyrme model:
\begin{eqnarray}
f_\pi^2&\equiv&\frac{4}{\pi}\k\Mkk^2=
\frac{1}{54\pi^4}\Mkk^2\lambda N_c\ , \label{fpi}
\end{eqnarray}
Another gauge we will consider below is
\begin{eqnarray}
\xi_+^{-1}(x^\mu)=\xi_-(x^\mu)=e^{i\Pi(x^\mu)/f_\pi}\ .
\end{eqnarray}
in terms of which the gauge fields are written as
\begin{eqnarray}
A_\mu(x^\mu,z) &=& \a_\m(x^\m) \wh{\psi}_0(z) + \b_\m(x^\m) +
\sum_{n=1}^\infty B_\mu^{(n)}(x^\mu)\psi_n(z)\ ,  \label{AzZeroGauge}  \\
\a_\m(x^\m) &=& \{\xi^{-1} , \dell_\mu \xi \} =
\frac{2i}{f_\pi}\dell_\mu \Pi  + [[\dell_\mu \Pi^3]] +
\calo(\Pi^4)\ , \nn  \\
\b_\m(x^\m) &=& \half[\xi^{-1} , \dell_\mu
\xi  ] =  \frac{1}{2 f_\pi^2}[\Pi, \dell_\mu \Pi] + \calo(\Pi^4)
\ , \nn
\end{eqnarray}
where  $[[\dell_\mu \Pi^3]] \equiv
-\frac{i}{3f_\pi^3}((\dell_\m\Pi) \Pi^2 + \Pi^2 \dell_\m \Pi -
2\Pi(\dell_\mu \Pi) \Pi ) $.

\section{Effective meson action: $n_B \neq 0$} \label{Newway}

We now extend the previous analysis to finite baryon density for
$n_B=0$. This is achieved by adding the background $U(1)_V$ field
$\Az$ to the fluctuating gauge field $A_M$.  Since, the vacuum
modes \{$\psi_n$, $\phi_n$\} are not mass eigenmodes in matter,
we may choose more pertinent eigenmodes in matter. Two basis set
are possible: (1) medium mass eigenmodes $\psi_n\sim
e^{-imt}f_n(z)$; (2) screening eigenmodes $\psi\sim
e^{i\vec{k}\cdot\vec{x}} f_n(z)$. With this in mind, we have the
following gauge fields decomposition
\begin{eqnarray}
A_0(x^\mu,z)&=& \Az(z) + \sum_{n=1}^\infty B_0^{(n)}(x^\mu)
\w_n(z)\ ,
\label{A0.2}\\
A_i(x^\mu,z)&=& \sum_{n=1}^\infty B_i^{(n)}(x^\mu)\psi_n(z)\ ,
\label{Ai.2}\\
A_z(x^\mu,z)&=& \sum_{n=0}^\infty \varphi^{(n)}(x^\mu)\phi_n(z)\, .
\label{Az.2}
\end{eqnarray}
$\Az(z)$ is the background gauge field. The time component modes
$(\w_n(z))$ and the space component $(\psi_n(z))$ are not necessarily
the same as Lorentz symmetry does not hold in the matter rest frame.
Note that $F_{\m z }$ is modified by $\Az$ while $F_{\m \n }$ is not.

In order to compute the DBI action (\ref{DBI.1}),
\begin{eqnarray}
&& S_{\DeDeB}^{DBI}
= -\wt T \int d^4x dz\, U^2  \nn \\
&&~~~~~~~~ \Tr \, \sqrt {\, 1 + (2\pi\alpha')^2 \frac{R^3}{2U^3}
 F_{\mu \nu}F^{\mu \nu}
+ (2\pi\alpha')^2 \frac{9}{4}\frac{U}{\Ukk}  F_{\mu z}F^{\mu z} +
[F^3] + [F^4] + [F^5]  } \nn \ ,
\end{eqnarray}
we need to know $ F_{\mu \n}F^{\mu \n}, F_{\mu z}F^{\mu z}, [F^3],
[F^4]$, and $[F^5]$, which are involved in general. To quadratic
order (ignoring $\calo((B_\m,\varphi)^3 )$), the contributions
are greatly simplified because of: 1) cyclic property of the trace,
2) antisymmetry of $F_{\m,\n}$, 3) parity of mode functions.
Then there is no contribution from $[F^3]$ and $[F^5]$. $[F^4]$ has
important terms that will modify $F_{\mu \n}F^{\mu \n}$:
\begin{eqnarray}
[F^4] = (2\pi\alpha')^4
\frac{9}{8}\frac{U}{\Ukk}\left(\frac{R}{U}\right)^3
F_{0z}F^{0z}F_{ij}F^{ij} + \calo((B_\m,\varphi)^4) \ .
\end{eqnarray}
\begin{table}[]
\begin{center}
\begin{tabular}{|c c l c c|}
\hline \vspace{-0.2cm}
& & & &  \\
\vspace{-0.2cm} $F_{\mu \nu}F^{\mu \nu}$ &$\ra$  & \Big[ $2\dell_0
B_i^{(n)} \dell^0 B^{(m)i} \psi_n\psi_m  + 2\dell_i B_0^{(n)}
\dell^{j}B^{(m)0}\omega_n\omega_m $  &  & \\
 & & & &  \\
 \vspace{-0.2cm} &  &  $- 2 \dell_0B_i^{(n)}
\dell^{i}B^{(m)0} \psi_n \psi_m$  &  &  \\
 & & & &  \\
$ $ & & $ +  (\dell_i B_j^{(n)}-\dell_{j}B_{i}^{(n)} )(\dell^i
B^{(m)j}-\dell^{j}B^{(m)i} )\psi_n\psi_m$ \Big] & $\equiv$ & $\a_2$\\
 & & & &  \\
\hline \vspace{-0.2cm}
& & & &  \\
$F_{\mu z}F^{\mu z} $ &$\ra$ & $ -(\dAz)^2  $ & $\equiv$ & $\b_0 $\\
\vspace{-0.2cm} & & & &  \\
$ $ & & $ +2 \dot{\cala}_0 \Big[\dell^0\varphi^{(n)} \phi_n -
B^{0(n)}\dot{\omega}_n + [B^{(n)0}, \varphi^{(m)}] \omega_n\phi_m
\Big] $ & $\equiv$ & $\b_1$ \\
\vspace{-0.2cm}& & & &  \\
$ $ & & $ + \Big[\dell_0 \varphi^{(n)} \dell^0 \varphi^{(m)}
\phi_n\phi_m  + B_0^{(n)}B^{(m)0}\dot{\omega}_n\dot{\omega}_m - 2
\dell_0
\varphi^{(n)} B^{(m)0}\phi_n \dot{\omega}_m $ & &  \\
\vspace{-0.2cm}& & & &  \\
$ $ & & $ + \dell_i \varphi^{(n)} \dell^i \varphi^{(m)}
\phi_n\phi_m  + B_i^{(n)}B^{(m)i}\dot{\psi}_n\dot{\psi}_m -
2\dell_i
\varphi^{(n)} B^{(m)i} \phi_n \dot{\psi}_m \Big]$ & $\equiv$ & $\b_2 $ \\
\vspace{-0.2cm}& & & &  \\
\hline %
\vspace{-0.2cm}& & & &  \\
$ [F^4] $ &$\ra$ & $ f_{ij}f^{ij}(\dAz)^2\psi_1^2 $
& $\equiv$ & $ \g_2 $\\
\vspace{-0.2cm} & & & &  \\
 \hline
\end{tabular}
\end{center}
  \caption{The relevant terms in evaluating the DBI action up to quadratic order in the
           fields ($B_\m, \varphi$). The upper dot stands for the derivative with respect
           to $z$. The terms should be understood in the integral and trace operation. }
  \label{FF}
\end{table}
Table (\ref{FF}) lists all the relevant terms, where we have introduced $f_{ij}$ defined as
\begin{eqnarray}
f_{ij} \equiv \dell_i v_j - \dell_j v_i \ ,
\end{eqnarray}
with $i,j = 1,2,3$. Table (\ref{FF}) should be understood in
the integral and trace operation. We omitted some terms
vanishing in those operations and rearranged some terms by using
the cyclicity of the trace.

In terms of the definitions on the RHS of the Table (\ref{FF}) ,
the action reads
\begin{eqnarray}
 S_{\DeDeB}^{DBI} = -\wt T \int d^4x dz\, U^2 \, \Tr \, \sqrt {\, P_0 +
P_1} \ ,
\end{eqnarray}
with
\begin{eqnarray}
P_0 &\equiv& 1 - (2\pi\alpha')^2 \frac{9}{4}\frac{U}{\Ukk} \ \b_0
= 1 -
  b K^{\frac{1}{3}}(\dZAz)^2  \ , \\
P_1 &\equiv&   (2\pi\alpha')^2 \frac{ R^3}{2 U^3}(\a_2) +
(2\pi\alpha')^2 \frac{9}{4}\frac{ U}{\Ukk} (\b_1+ \b_2 ) \nn \\
&+&(2\pi\alpha')^4 \frac{9}{8} \frac{ R^3}{U_{KK}U^2 } (\g_2) \ ,
\label{P1}
\end{eqnarray}
where $P_0$ does not contain meson fields but involves the
baryon density. Expanding the action for small fields we have
\begin{eqnarray}
S_{\DeDeB}^{DBI} &=& -\wt T \int d^4x dz\, U^2 \,\Tr \, \left[
\sqrt {\, P_0} + \half \frac{P_1}{\sqrt{P_0}}
-\frac{1}{8}\frac{P_1^2}{{\sqrt{P_0}}^3}  \right] + \calo((B_\m, \varphi)^3) \nn \\
&=& S_1 + S_2 +  \calo((B_\m, \varphi)^3) \ ,
\end{eqnarray}
with
\begin{eqnarray}
&& S_1 \equiv -\wt T \int d^4x dz\, U^2 \Tr \, \D^{-1} \ , \\
&& S_2 \equiv -\wt T \int d^4x dz\, U^2 \Tr \, \left[\half \D P_1
 - \frac{1}{8} \D^3 P_1^2 \right]\  ,  \label{S2}
\end{eqnarray}
where the modification factor $\D(n_B)$ is
\begin{eqnarray}
\D(n_B) \equiv  \frac{1}{\sqrt{P_0}} = \frac{1}{\sqrt {\, 1 -
  b K^{\frac{1}{3}}(\dZAz)^2} } = \sqrt{1+\frac{n_B^2} {4 a^2 b} K^{-5/3} } \ . \nn
\end{eqnarray}
$-S_0$ is the grand potential discussed in section
\ref{Sec:Thermodynamics}, and $S_2$ will be reduced to
\footnote{Note that the pattern: $\D$, $\D^{-1}$, and $\D^{3}$.
This pattern appears also when we consider higher order terms including couplings. The origin is
explained in  Appendix A. }
\begin{eqnarray}
&& S_2 = -\Tr \int d^4 x \Bigg\{ \nn \\
&&\quad \ \ \ \left[\int dZ K^{-1/3}\D \ \Psi_n \Psi_m\right]
\dell_0 B_i^{(m)} \dell^0 B^{(n)i}
+ \left[\int dZ K^{-1/3}\D\ \Omega_n \ \Omega_m\right]
\dell_i B_0^{(n)} \dell^i B^{(m)0}  \nn \\
&&\quad  - \left[\int dZ K^{-1/3}\D\ \Psi_n \ \Omega_m\right]
2 \dell_0 B_i^{(n)} \dell^i B^{(m)0} \nn \\
&& \quad  + \left[\int dZ K^{-1/3}\D^{-1}\Psi_n\Psi_m\right]\half
(\dell_i B_j^{(n)}-\dell_{j}B_{i}^{(n)} )(\dell^i
B^{(m)j}-\dell^{j}B^{(m)i} ) \nn \\
&&\quad  + \left[\Mkk^2 \int dZ K \D^3\ \dell_Z\Omega_n\ \dell_Z
\Omega_m\right] B_0^{(n)} B^{(m)0}
 + \left[\Mkk^2 \int dZ K \D \ \dell_Z\Psi_n
\dell_Z\Psi_m \right] B_i^{(n)}B^{(m)i} \nn \\
&&\quad + \left[ \Mkk^2 \int dZ K \D^3  \Phi_n \Phi_m \right]
\dell_0\varphi^{(n)}\dell^0\varphi^{(m)} + \left[\Mkk^2\int dZ K
\D  \Phi_n \Phi_m \right]
\dell_i\varphi^{(n)}\dell^i\varphi^{(m)} \nn
\\
&&\quad - \left[\Mkk^2\int dZ K \D^3  \Phi_n \dell_Z \Omega_m
\right] 2 \dell_0\varphi^{(n)} B^{(m)0}  - \left[\Mkk^2\int dZ K
\D  \Phi_n \dell_Z \Psi_m \right] 2 \dell_i\varphi^{(n)}
B^{(m)i}\Bigg\}  \ ,  \nn \\
\end{eqnarray}
where we defined the scaled eigenfounctions as
\begin{eqnarray}
\Omega_n \equiv \sqrt{\k} \w_n \ , \quad \Phi_n \equiv \sqrt{\k}
\psi_n \ , \quad \Phi_n \equiv \sqrt{\k} \Ukk \phi_n \ .
\end{eqnarray}
At zero density $\D = 1$, so $\Phi_n = \Omega_n$ and the action
reduces to the (\ref{SSaction})  by the same mode function in
(\ref{psi}) $\sim$ (\ref{phi:ortho}). However at finite density
the eigen modes $\Omega_{n}$, $\Psi_{n}$, and $\Phi_n$ cannot be
determined uniquely. In other words there is no mode
decomposition which makes the action completly diagonal. So we
consider the space-like and time-like separatly: (1) $A_M =
A_M(x^i,z)$ and (2) $A_M = A_M(x^0,z)$.

\subsection{Space-like fields $A_M = A_M(x^i,z)$}

First we consider time-independent gauge fields. Up to quadratic
order the action is
\begin{eqnarray}
&& S_2 = -\Tr \int d^4 x \Bigg\{ \left[\int dZ K^{-1/3}\D\
\Omega_n \ \Omega_m\right] \dell_i B_0^{(n)} \dell^i B^{(m)0} \nn
\\ && \quad  + \left[\int dZ
K^{-1/3}\D^{-1}\Psi_n^S\Psi_m^S\right]\half (\dell_i
B_j^{(n)}-\dell_{j}B_{i}^{(n)} )(\dell^i
B^{(m)j}-\dell^{j}B^{(m)i} ) \nn \\
&&\quad +  \left[\Mkk^2 \int dZ  K \D^3\ \dell_Z\Omega_n\ \dell_Z
\Omega_m\right]  B_0^{(n)} B^{(m)0} + \left[\Mkk^2 \int dZ
K \D \ \dell_Z\Psi_n^S  \dell_Z\Psi_m^S \right] B_i^{(n)}B^{(m)i} \nn \\
&&\quad + \left[\Mkk^2\int dZ K \D  \Phi_n^S \Phi_m^S \right]
\dell_i\varphi^{(n)}\dell^i\varphi^{(m)} - \left[\Mkk^2\int dZ K
\D  \Phi_n^S \dell_Z \Psi_m^S \right] 2 \dell_i\varphi^{(n)}
B^{i(m)} \Bigg\} \ , \nn \\
\end{eqnarray}
where we have defined the scaled eigenfunctions as
\begin{eqnarray}
\Omega_n \equiv \sqrt{\k} \w_n \ , \quad \Psi_n^S \equiv \sqrt{\k}
\psi_n \ , \quad \Phi_n^S \equiv \sqrt{\k} \Ukk \phi_n \ .
\end{eqnarray}
To diagonalize the action we choose $\Psi_n^S$ as the
eigenfunction satisfying
\begin{eqnarray}
&&-K^{1/3} \D^{-1}  \dell_Z\left( K \D^3 \ \dell_Z\Omega_n\right)=
\lambda_n^\Omega\Omega_n \ , \\
&&-K^{1/3} \D \ \dell_Z\left( K \D \ \dell_Z\Psi_n^S\right)=
\lambda_n^S\Psi_n^S \ ,
\end{eqnarray}
with the normalization conditions,
\begin{eqnarray}
&&\int dZ K^{-1/3}\D\ \Omega_n \ \Omega_m = \d_{nm} \ , \\
&&\int dZ K^{-1/3}\D^{-1}\Psi_n^S\Psi_m^S = \d_{nm} \ ,
\label{PsiSEqn}
\end{eqnarray}
which imply
\begin{eqnarray}
&& \int dZ  K \D^3\ \dell_Z\Omega_n\ \dell_Z \Omega_m = \lambda_n^\Omega \d_{nm} \ , \\
&& \int dZ K \D \ \dell_Z\Psi_n^S  \dell_Z\Psi_m^S = \lambda_n^S
\d_{nm} \ .
\end{eqnarray}
If we choose $\Phi_n^S$  as
\begin{eqnarray}
\Phi_n^S = \inv{\Mkk \sqrt{\l_n^S}} \dell_Z \Psi_n^S\ \ (n\ge1) \
, \quad  \Phi_0^S = \inv{\Mkk}\inv{\sqrt{\int dZ(K^{-1}
\D^{-1})}}\ \inv {K \D } \ ,
\end{eqnarray}
then $\dell_i\varphi^{(n)} $ ($n\ge1$)  can be absorbed into
$B_i^{(n)}$ through the gauge transformation
\begin{eqnarray}
B_i^{(n)} \ra B_i^{(n)} + \inv{\Mkk \sqrt{\l_n^S}}
\dell_i\varphi^{(n)} \ .
\end{eqnarray}
These choices of mode functions reduces the action to
\begin{eqnarray}
&&S_2 = -\Tr \int d^4 x \Big\{
\dell_i\varphi^{(0)}\dell^i\varphi^{(0)} \nn \\
&& \qquad \qquad \qquad + \dell_i B_0^{(n)} \dell^i B^{(n)0} +
\half f_{ij}^{(n)}f^{(n)ij} + \calm_{n}^{\shortparallel\, 2}
B_0^{(n)} B^{(n)0} + \calm_{n}^{\perp\, 2} B_i^{(n)}B^{(n)i}
\Big\} \ ,
\end{eqnarray}
where we have defined longitudinal screening masses
$\calm_{n}^\shortparallel$ and transverse screening masses
$\calm_{n}^\perp$ as
\begin{eqnarray}
\calm_{n}^\shortparallel \equiv \sqrt{\l_n^\Omega}\Mkk \ , \qquad
\calm_{n}^\perp \equiv \sqrt{\l_n^S}\Mkk \ . \label{MassDef.1}
\end{eqnarray}
\subsection{Time-like fields $A_M=A_M(x^0,z)$ }
For spacially homogeneous gauge fields the action reads
\begin{eqnarray}
&& S_2 = -\Tr \int d^4 x \Bigg\{ \left[\int dZ K^{-1/3}\D \
\Psi_n^T \Psi_m^T \right]  \dell_0 B_i^{(m)}
\dell^0 B^{(n)i}  \nn \\
&&\quad  + \left[\Mkk^2 \int dZ K \D \ \dell_Z\Psi_n^T
\dell_Z\Psi_m^T \right] B_i^{(n)}B^{(m)i}+ \left[\Mkk^2 \int dZ K
\D^3\ \dell_Z\Omega_n\ \dell_Z \Omega_m\right] B_0^{(n)}
B^{(m)0} \nn \\
&&\quad + \left[ \Mkk^2 \int dZ K \D^3  \Phi_n^\Omega
\Phi_m^\Omega \right] \dell_0\varphi^{(n)}\dell^0\varphi^{(m)} -
\left[\Mkk^2\int dZ K \D^3  \Phi_n^\Omega \dell_Z \Omega_m \right]
2 \dell_0\varphi^{(n)} B^{0(m)} \Bigg\}  \ , \nn \\
\end{eqnarray}
where we have defined the scaled eigenfunctions
\begin{eqnarray}
\Omega_n \equiv \sqrt{\k} \w_n \ , \quad \Psi_n^T \equiv \sqrt{\k}
\psi_n \ , \quad \Phi_n^\Omega \equiv \sqrt{\k} \Ukk \phi_n \ .
\end{eqnarray}
We choose $\Psi_n^S$ as the eigenfunction satisfying
\begin{eqnarray}
&&-K^{1/3} \D^{-1} \dell_Z\left( K \D \ \dell_Z\Psi_n^T \right)=
\lambda_n^T\Psi_n^T \ , \label{PsiTEqn}\\
&&-K^{1/3} \D^{-1}  \dell_Z\left( K \D^3 \ \dell_Z\Omega_n
\right)= \lambda_n^\Omega\Omega_n \ , \label{OmegaEqn}
\end{eqnarray}
with the normalization conditions,
\begin{eqnarray}
&&\int dZ K^{-1/3}\D \ \Psi_n^T \Psi_m^T  = \d_{nm} \ , \\
&&\int dZ K^{-1/3}\D\ \Omega_n \ \Omega_m = \d_{nm} \ ,
\end{eqnarray}
which imply
\begin{eqnarray}
&& \int dZ K \D \ \dell_Z\Psi_n^T \dell_Z\Psi_m^T  = \lambda_n^T
d_{nm}
\ , \\
&& \int dZ  K \D^3\ \dell_Z\Omega_n\ \dell_Z \Omega_m =
\lambda_n^\Omega \d_{nm} \ ,
\end{eqnarray}
If we choose $\Phi_n^S$  as
\begin{eqnarray}
\Phi_n^\Omega = \inv{\Mkk \sqrt{\l_n^\Omega}} \dell_Z
\Psi_n^\Omega \ , \quad \Phi_0^\Omega = \inv{\Mkk}\inv{\sqrt{\int
dZ(K^{-1} \D^{-3})}}\ \inv {K \D^{3} } \ ,
\end{eqnarray}
then $\dell_0\varphi^{(n)} $ ($n\ge1$)  can be absorbed into
$B_0^{(n)}$ through the gauge transformation
\begin{eqnarray}
 B_0^{(n)} \ra B_0^{(n)} + \inv{\Mkk
\sqrt{\l_n^\Omega}} \dell_0\varphi^{(n)} \ .
\end{eqnarray}
The action is reduced to
\begin{eqnarray}
S_2 = -\Tr \int d^4 x
\Big\{\dell_0\varphi^{(n)}\dell^0\varphi^{(n)}+ \dell_0 B_i^{(n)}
\dell^0 B^{(n)i} + m_n^2 B_i^{(n)}B^{(n)i}+ \Mkk^2 \l_n^\Omega
B_0^{(n)} B^{(n)0}
 \Big\} \ ,
\end{eqnarray}
where we have defined the mass
\begin{eqnarray}
m_n = \sqrt{\l_n^T}\Mkk \ . \label{MassDef.2}
\end{eqnarray}

\subsection{Pion effective action} \label{Skyrme.Action}
In this subsection we work in the $A_z = 0$ gauge following the
procedure in section \ref{A_z=0.gauge}.
\subsubsection{Time-like field ($A_M=A_M(x^0,z)$) }
First consider the case $A_M = A_M(x^0,z)$. By the gauge
transformation $A_M\ra g A_M g^{-1}+ g\dell_M g^{-1}$ with the
gauge function
\begin{eqnarray}
g^{-1}(x^0,z)= P \exp\left\{- \int_{0}^z dz'\, A_z(x^0,z')
\right\} \ , \label{ginv}
\end{eqnarray}
the gauge fields are rewritten as
\begin{eqnarray}
A_0(x^0,z) &=& \cala_0(z) + \xi_+(x^0)\dell_0
\xi_+^{-1}(x^0)\omega_+(z)+
\xi_-(x^0)\dell_0 \xi_-^{-1}(x^0)\omega_-(z)  \ , \\
A_i(x^0,z)&=& A_z(x^0,z)= 0 \ ,
\end{eqnarray}
where we have omitted the vector mesons $B_\m^{(n)}$. The $\w_\pm$ are
obtained as zero mode solutions of (\ref{OmegaEqn}) satisfying
the boundary condition for $A_0(x^{0},z)$:
\begin{eqnarray}
\omega_\pm(z) \equiv \half \pm \inv{\int dZ(K^{-1} \D^{-3})}\
\int_{0}^{Z} dZ \inv {K \D^{3} } \ .
\end{eqnarray}
By using the residual gauge symmetry $h(x^\m)$
(\ref{Hidden.gauge.1}) and (\ref{Gauge.field.for.Skyrme}) we may
express the gauge field as
\begin{eqnarray}
A_0(x^0,z) = \cala_0 + U^{-1}(x^0)\dell_0 U(x^0)\omega_+(z) \ .
\end{eqnarray}
The field strength is
\begin{eqnarray}
F_{z\m}= \dAz +  U^{-1}\dell_0 U \wh{\phi}_0^\omega(z) \ , \qquad
F_{\m\n} = 0\ ,
\end{eqnarray}
where
\begin{eqnarray}
\wh{\phi}_0^\omega(z) \equiv  \dell_z\omega_+(z) = \inv{\Ukk\int
dZ(K^{-1} \D^{-3})}\ \inv {K \D^{3} } \ .
\end{eqnarray}
The action becomes
\begin{eqnarray}
S_2 = \Tr \int d^4 x  \left[\k \Mkk^2 \inv{\int dZ K^{-1}\D^{-3}}
\right] (U^{-1}\dell_0 U)^2 \ ,
\end{eqnarray}
and we identify the time-like pion decay constant ${f^{T}_\pi} $  as
\begin{eqnarray}
{f^{T}_\pi}^2 = \frac{4 \k \Mkk^2}{\int dZ K^{-1}\D^{-3}} \ ,
\label{fpiT}
\end{eqnarray}
by comparison with the Skyrme model.

\subsubsection{Space-like field ($A_M=A_M(x^i,z)$) }

Similarly, we consider the case $A_M = A_M(x^i,z)$. Using
the same gauge transformation we can work with the gauge fields,
\begin{eqnarray}
A_i(x^i,z) &=& \xi_+(x^i)\dell_i \xi_+^{-1}(x^i)\psi^S_+(z)+
\xi_-(x^i)\dell_i \xi_-^{-1}(x^i)\psi^S_-(z) \ , \nn \\
A_0(x^i,z) &=& \Az(z) \ , \qquad   A_z(x^i,z) = 0 \ , \nn
\end{eqnarray}
where $\psi^S_\pm$ are obtained as a zero mode solution of
(\ref{PsiSEqn}) satisfying the pertinent boundary condition of
$A_i(x^{i},z)$:
\begin{eqnarray}
\psi^S_\pm(z) \equiv \half \pm \inv{\int dZ(K^{-1} \D^{-1})}\
\int_{0}^{Z} dZ \inv {K \D } \label{psiShat}
\end{eqnarray}
Then the gauge field and the field strength in the gauge
(\ref{Hidden.gauge.1}) are
\begin{eqnarray}
A_0(x^0,z) &=& \cala_0 + U^{-1}(x^i)\dell_i U(x^i)\psi^S_+(z)
\nn \\
F_{z\m} &=& \dAz +  U^{-1}\dell_i U \wh{\phi}_0^S (z) \ ,
\end{eqnarray}
where we do not consider $F_{\m\n}$ since we are interested in
the kinetic part and
\begin{eqnarray}
\wh{\phi}^S_0(z) \equiv  \dell_z\psi^S_+(z) = \inv{\Ukk\int
dZ(K^{-1} \D^{-1})}\ \inv {K \D } \ .
\end{eqnarray}
The action is
\begin{eqnarray}
&& S_2 = \Tr \int d^4 x \left[\k \Mkk^2 \inv{\int dZ
K^{-1}\D^{-1}} \right] (U^{-1}\dell_i U)^2 \ ,
\end{eqnarray}
and ${f^{S}_\pi}$ is identified by
\begin{eqnarray}
{f^{S}_\pi}^2 = \frac{4 \k \Mkk^2}{\int dZ K^{-1}\D^{-1}} \ .
\label{fpiS}
\end{eqnarray}

\subsection{Vector Mesons Interactions}
In this section we study the interactions of the fields
$B_0^{(1)}$, $B_i^{(1)}$ and $\varphi^{(0)}$ corresponding to the
lowest medium modes $\Omega_1$, $\Psi_1$, and $\Phi_1$. For
simplicity, we use the following notation,
\begin{eqnarray}
v_0 \equiv B_0^{(1)}\ , \quad v_i \equiv B_i^{(1)}\ , \quad \Pi
\equiv \varphi^{(0)} \ .
\end{eqnarray}
The details of the computation are relegated to Appendix
A~\footnote{Both in Appendix A and this section, the vector meson
field is considered as anti Hermitian. Although in Appendix A, we
are working with the vacuum modes instead of the medium modes, the
conversion can be done by inspection using the formula tabulated
in Table (\ref{t2},\ref{t4})}.
\subsubsection{Time-like Fields $A_M=A_M(x^0,z)$ }
\begin{eqnarray}
&&S_2 = \Tr \int d^4 x \Big\{-\dell_0\Pi \dell^0\Pi+ \dell_0 v_i
\dell^0 v^i + m_1^2 v_i v^i+ \Mkk^2 \l_1^\Omega v_0  v^0
 \Big\}   \nn \\
&& \qquad \qquad \qquad -2g^T_{v\Pi^2}\, v_0 [\Pi , \dell^0 \Pi ]
+ g_{v^3}^T \, 2 \dell_0 v_i [v^0 ,v^i] + \cdots \Big\} \ ,
\end{eqnarray}
where the couplings can be read from~(\ref{t2},\ref{t4}) in
Appendix A by substituting the vacuum mode functions by the medium
mode functions
\begin{eqnarray}
&&g_{v\Pi^2}^T = \inv{\sqrt{\k}}  \displaystyle{\frac{\int dZ\,
\frac{\Omega_1}{K\D^3}}{\int dZ\, \frac{1}{K\D^3}}} \ , \nn \\
&&g_{v^3}^T = \inv{\sqrt{\k}} \int dZ\, K^{-1/3} \Omega_1
(\Psi_1^T)^2  \D \ . \label{Coupling.T}
\end{eqnarray}

\subsubsection{Space-like Fields $A_M=A_M(x^i,z)$}
\begin{eqnarray}
&& S_2 = \Tr \int d^4 x \Big\{ -\dell_i\Pi \dell^i\Pi+ \dell_i v_0
\dell^i v^0 + \half f_{ij} f^{ij} + \calm_{1}^{\shortparallel\,
2} v_0 v^0 + \calm_{1}^{\perp\, 2} v_i v^i  \nn \\
&& \qquad \qquad \qquad -2g^S_{v\Pi^2}\, v_i [\Pi , \dell^i \Pi ]
+ g_{v^3}^S \, 2 \dell_iv_0 [v^i ,v^0] +  \widetilde{g}_{v^3}^S\,
f_{ij}[v^i ,v^j] + \cdots \Big\} \ ,
\end{eqnarray}
where the couplings can be read from (\ref{t2},\ref{t4}) in
Appendix A, again  by substituting the vacuum  mode functions by
the medium mode functions

\begin{eqnarray}
&&g_{v\Pi^2}^S = \inv{\sqrt{\k}}  \displaystyle{\frac{\int dZ\,
\frac{\Psi^S_1}{K\D}}{\int dZ\, \frac{1}{K\D}}} \ , \nn \\
&&g_{v^3}^S = \inv{\sqrt{\k}} \int dZ\, K^{-1/3} (\Omega_1)^2
\Psi_1^S  \D^{-1} \ , \nn \\
&&\widetilde{g}_{v^3}^S = \inv{\sqrt{\k}} \int dZ\, K^{-1/3}
(\Psi_1^S)^3  \D^{-1} \ . \label{Coupling.S}
\end{eqnarray}
\subsubsection{Zero Density Limit}
To check the current mode decomposition used in this section,
we take the zero baryon density limit. In this case, Lorentz
symmetry is enforced and the action reads
\begin{eqnarray}
&&S_2 = \Tr \int d^4 x \Big\{-\dell_\m\Pi \dell^\n\Pi+ \half
f_{\m\n} f^{\m\n} + m_1^2 v_\m v^\n
 \Big\}    \nn \\
&& \qquad \qquad \qquad -2g_{v\Pi^2}\, v_\m [\Pi , \dell^\m \Pi ]
+ g_{v^3} \,  f_{\m\n} [v^\m ,v^\n] + \cdots \Big\} \ ,
\label{Action.zero.density}
\end{eqnarray}
which is the same as Eqn.(5.40) in \cite{Sakai1} except the
$v\Pi\Pi$ coupling. The difference comes from the gauge choice.
In \cite{Sakai1} $A_z = 0$ gauge is used and we chose $A_M(z \ra
\infty) \ra 0 $.  Since the difference is merely a gauge choice,
physics will not be changed. However we will repeat the analysis
of couplings at zero density with (\ref{Action.zero.density}),
since the action in our gauge is more convenient for reading off
physical quantities. Also it is readily extendable to finite
baryon density.

First we examine the KSRF relation by defining $a_\textrm{KSRF}$
as
\begin{eqnarray}
a_\textrm{KSRF} \equiv \frac{4\, g_{v\Pi^2}^2\, {f_\pi^2}}{m_1^2 }
\ \sim \  \left\{
       \begin{array}{ll}
          2.03 & \textrm{Experiment} \\
          1.3 & \textrm{Sakai Sugimoto model}
       \end{array} \right. \ ,
\end{eqnarray}
which is the same value reported in \cite{Sakai1, Sakai2}, as
expected. The universality of the vector meson coupling can be
checked by $a_\textrm{U}$ defined as
\begin{eqnarray}
a_\textrm{U} \equiv  \frac{g_{v\Pi^2}}{g_{v^3}} \ \sim \  \left\{
       \begin{array}{ll}
          1 & \textrm{The universality of the vector meson coupling} \\
          0.93 & \textrm{Sakai Sugimoto model}
       \end{array} \right. \ ,
\end{eqnarray}
which is also the same value as in~\cite{Sakai1}. Notice that both
relations include $g_{v\Pi^2}$ and can be read from
(\ref{Action.zero.density}). In the $A_z = 0$ gauge we
should convert $g_{v\Pi^2}$ to $a_{v\Pi^2}$ by~\cite{Sakai1}
\begin{eqnarray}
a_{v\Pi^2} = \frac{2 g_{v\Pi^2} }{m_1^2} \ . \label{atog}
\end{eqnarray}
When we consider the field redefinition in
(\ref{Action.zero.density})
\begin{eqnarray}
v_\m \ra v_\m + \frac{a_{v^3}}{2}[\Pi, \dell_\m \Pi] \ ,
\end{eqnarray}
the algebraic relation (\ref{atog}) appears immediate. However when we
look at the integral expression of $g_{v\Pi^2}$ and $a_{v\Pi^2}$
the equivalence is obscured.
\begin{eqnarray}
&& \quad a_{v\Pi^2} = \frac{2 g_{v\Pi^2} }{m_1^2} \nn \\
&& \Leftrightarrow  \frac{\pi^2}{8}\int dZ K^{-1/3}\Psi_1 (1-4
\widehat{\psi}_0^2) \int dZ K (\dell_Z \Psi_1)^2 = \int dZ K^{-1}
\Psi_1 \ .
\end{eqnarray}

Next and following~\cite{Sakai1}, we compare
(\ref{Action.zero.density}) with the action from the hidden local
symmetry approach
\begin{eqnarray}
&&S_H \equiv \Tr \int d^4 x \Big\{-\dell_\m\Pi \dell^\n\Pi+ \half
f_{\m\n} f^{\m\n} + a g^2 f_\pi^2 v_\m v^\n
 \Big\}  \nn \\
&& \qquad \qquad \qquad -a g\, v_\m [\Pi , \dell^\m \Pi ] + g \,
f_{\m\n} [v^\m ,v^\n] + \cdots \Big\} \ .
\end{eqnarray}
The hidden local symmetry parameter (LHS) can be written in terms
of the D-brane effective action parameter (RHS):
\begin{eqnarray}
&& g = g_{v^3} \ ,  \\
&& a = \frac{2 g_{v\Pi^2}}{g} = \frac{2 g_{v\Pi^2}}{g_{v^3}} \ , \label{Hidden2} \\
&& f_\pi^2 = \frac{m_1^2}{a g^2} = \frac{m_1^2}{2 g_{v^3}
\label{Hidden3} g_{v\Pi^2}} \ ,
\end{eqnarray}
where we used the first two relations to get the last. We may
define the parameter $a_\textrm{H}$ which quantify the difference
between hidden local symmetry approach and our model:
\footnote{For a=2 the hidden local symmetry approach implies KSRF
relation and the universality of the vector meson coupling.
Here, we do not require this value since we want to compare our
model with the hidden local symmetry itself.}
\begin{eqnarray}
a_\textrm{H} \equiv \frac{ 2 g_{v^3} g_{v\Pi^2}{f_\pi^2}}{m_1^2}
\ \sim \  \left\{
       \begin{array}{ll}
          1 & \textrm{Hidden local symmetry} \\
          0.72 & \textrm{Sakai Sugimoto model}
       \end{array} \right. \ ,
\end{eqnarray}
which is the same value reported in \cite{Sakai1}, as expected.
$a_\textrm{H}$ may be interpreted as follows. Since $f_\pi$ is an
input parameter the Hidden local symmetry has two adjustable
parameters, so $a$ is not uniquely determined. It can be fixed by
(\ref{Hidden2}) or (\ref{Hidden3}). When these two procedures
yield the same value, $a_\textrm{U}  = 1$.
\subsection{Numerical results}
All the numerical work reported here has been carried out for the
lowest modes $v_\m \equiv B_\m^{(1)}$ and $\Pi \equiv \varphi^{(0)}$
with the parameters discussed in section 3.
\subsubsection{Mass and Screening Mass}
\begin{figure}[!ht]
  \begin{center}
  \subfigure[] {\includegraphics[width=8cm]{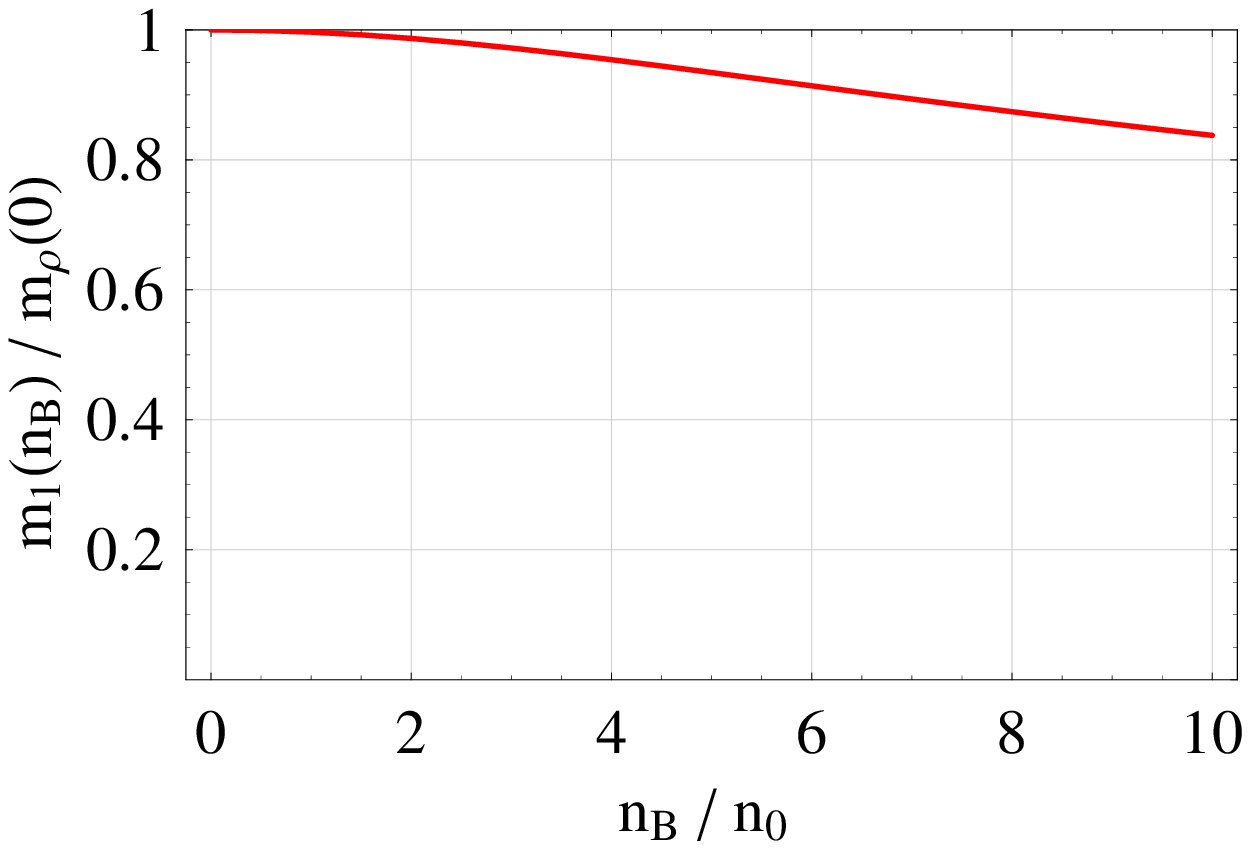}}
  \subfigure[] {\includegraphics[width=8cm]{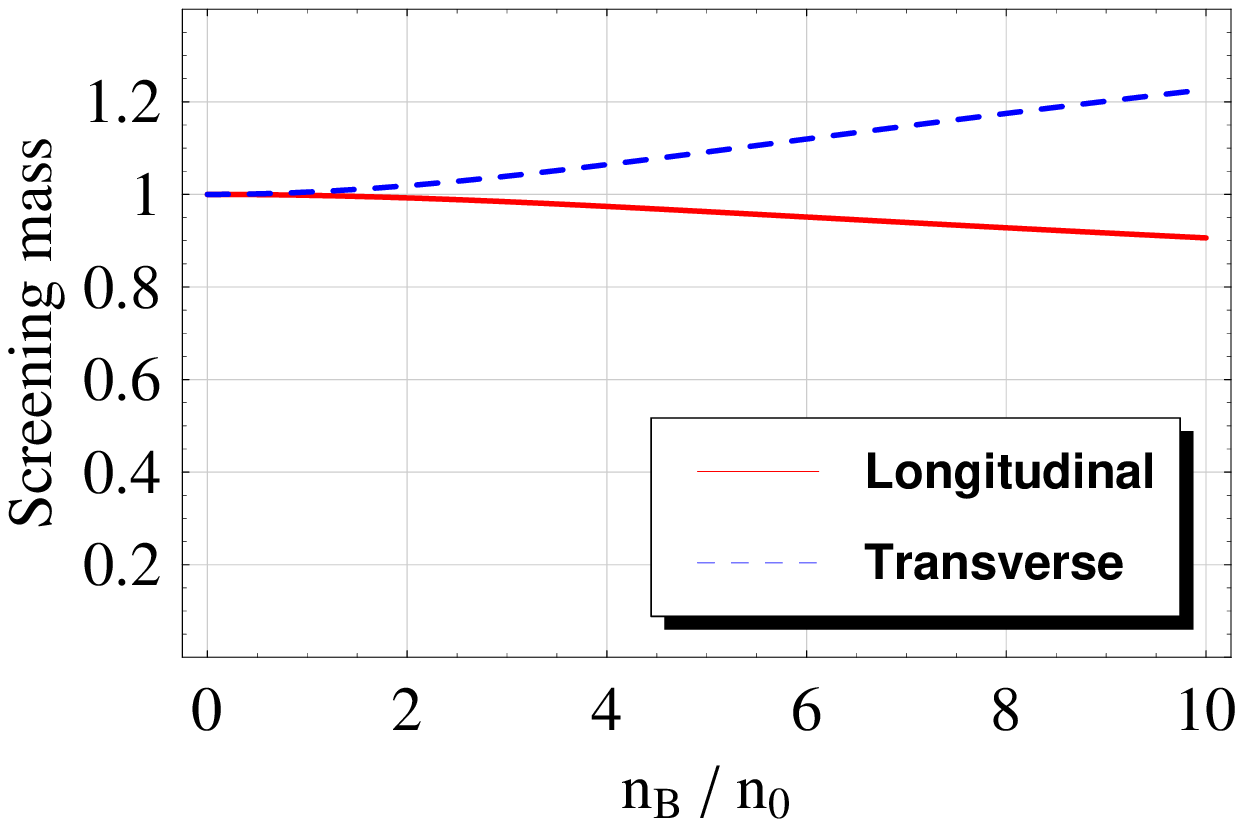}}
  \caption{(a) Mass (b) Screening mass (Longitudinal mode: $\calm_{1}^\shortparallel(n_B) / m_\rho(0) $,
   Transverse mode: $\calm_{1}^\perp(n_B) /m_\rho(0)$)}
  \label{Fig:mass}
  \end{center}
\end{figure}

From the previous section the meson masses (\ref{MassDef.2})
(time-like) and the screening masses (\ref{MassDef.1})
(space-like) are defined as
\begin{eqnarray}
&&m_n \equiv \sqrt{\l_n^T}\Mkk \ , \nn \\
&& \calm_{n}^\shortparallel \equiv \sqrt{\l_n^\Omega}\Mkk \ ,
\qquad \calm_{n}^\perp \equiv \sqrt{\l_n^S}\Mkk \ ,
\end{eqnarray}
where  $\l_n^T$, $\l_n^\Omega$, and $\l_n^S$ are determined as the
eigenvalues of the following equations
((\ref{PsiTEqn}),(\ref{OmegaEqn}),(\ref{PsiSEqn})), respectively:
\begin{eqnarray}
&&-K^{1/3} \D^{-1} \dell_Z\left( K \D \ \dell_Z\Psi_n^T \right)=
\lambda_n^T\Psi_n^T \ , \\
&&-K^{1/3} \D^{-1}  \dell_Z\left( K \D^3 \ \dell_Z\Omega_n\right)=
\lambda_n^\Omega\Omega_n \ , \\
&&-K^{1/3} \D \ \dell_Z\left( K \D \ \dell_Z\Psi_n^S\right)=
\lambda_n^S\Psi_n^S \ .
\end{eqnarray}
Their dependense on the baryon density normalized to the nuclear
matter density is shown in~(\ref{Fig:mass}) for  the lowest
eigenmode. The time-like and transverse screening mass are seen
to decrease midly with density. The longitudinal screening mass
increases moderatly with baryon density. The mild dependence on
the density for the SS model indicates that the vector mesons are
weakly affected by the baryon density in this version of the SS
model. As the inserted baryons are point like, at large $N_c$
their interaction is chiefly repulsive through $\omega$'s as
induced by D8-$\DeB$. The $\omega$ interactions with vectors and
axials is mostly anomalous (through the WZ term) and therefore
small as we ignored the WZ term.

\subsubsection{Pion decay constant}
\begin{figure}[!hb]
  \begin{center}
  \includegraphics[width=8cm]{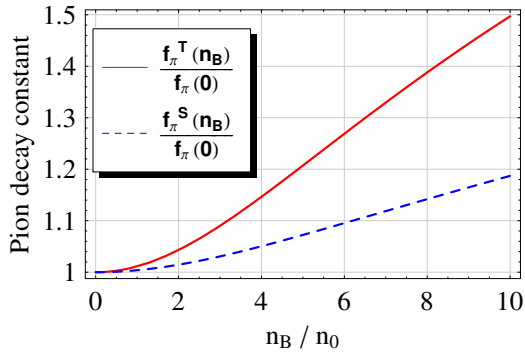}
  \caption{Pion decay constant}
  \label{Fig:pion}
  \end{center}
\end{figure}

The pion decay constant is identified from ((\ref{fpiT}),(\ref{fpiS}))
respectively,
\begin{eqnarray}
&&{f^{T}_\pi}^2 = \frac{4 \k \Mkk^2}{\int dZ K^{-1}\D^{-3}}\ , \nn \\
&&{f^{S}_\pi}^2 = \frac{4 \k \Mkk^2}{\int dZ K^{-1}\D^{-1}}\ .
\end{eqnarray}
The explicit dependence on the baryon density is shown in~Fig.(\ref{Fig:pion}).
Both the time-like and space-like pion decay constant are found
to {\it increase} with the baryon density. The increase is
quadratic at small densities. Since the S-wave pion scattering
length with baryons is $1/N_c$ this explains the absence of a
linear term. Moreover, for point-like external baryon sources the
pion-Axial-Vector coupling in matter at the origin of the pion
decay constant involves two baryon sources and is repulsive.
\subsubsection{Vector Couplings and KSRF Relation}
The vector couplings are identified in (\ref{Coupling.T}) and
(\ref{Coupling.S}). Their overall dependence on the baryon
density is again mild as explained above.
\\
\\
{\bf $\mathbf{v \Pi \Pi}$ couplings}:
\begin{eqnarray}
&&g_{v\Pi^2}^T = \inv{\sqrt{\k}}  \displaystyle{\frac{\int dZ\,
\frac{\Omega_1}{K\D^3}}{\int dZ\, \frac{1}{K\D^3}}} \ , \nn \\
&&g_{v\Pi^2}^S = \inv{\sqrt{\k}}  \displaystyle{\frac{\int dZ\,
\frac{\Psi^S_1}{K\D}}{\int dZ\, \frac{1}{K\D}}} \ , \nn \\
\end{eqnarray}
{\bf $\mathbf{v v v}$ couplings}:
\begin{eqnarray}
&&g_{v^3}^T = \inv{\sqrt{\k}} \int dZ\, K^{-1/3} \Omega_1
(\Psi_1^T)^2  \D \ \nn \\
&&g_{v^3}^S = \inv{\sqrt{\k}} \int dZ\, K^{-1/3} (\Omega_1)^2
\Psi_1^S  \D^{-1} \ , \nn \\
&&\widetilde{g}_{v^3}^S = \inv{\sqrt{\k}} \int dZ\, K^{-1/3}
(\Psi_1^S)^3  \D^{-1} \ .
\end{eqnarray}
\begin{figure}[!hb]
  \begin{center}
  \subfigure[] {\includegraphics[width=8cm]{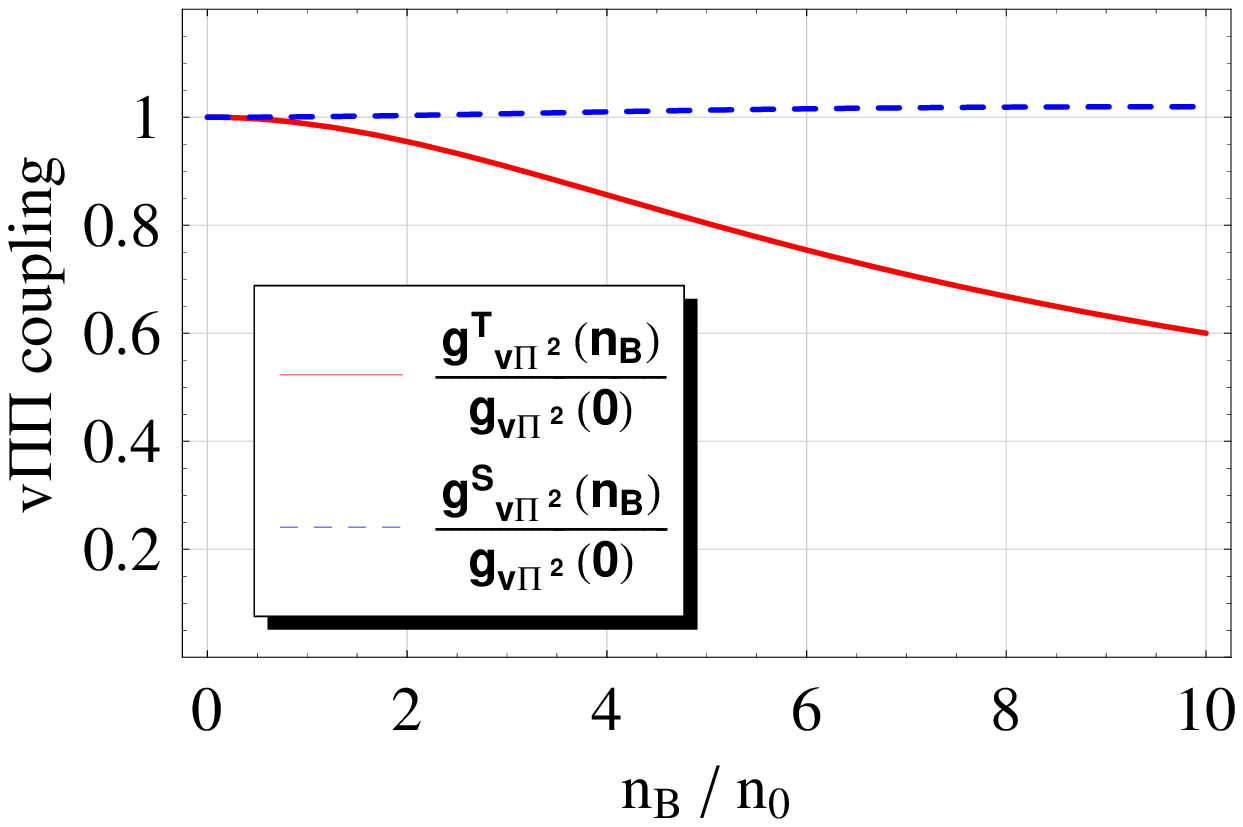}}
  \subfigure[] {\includegraphics[width=8cm]{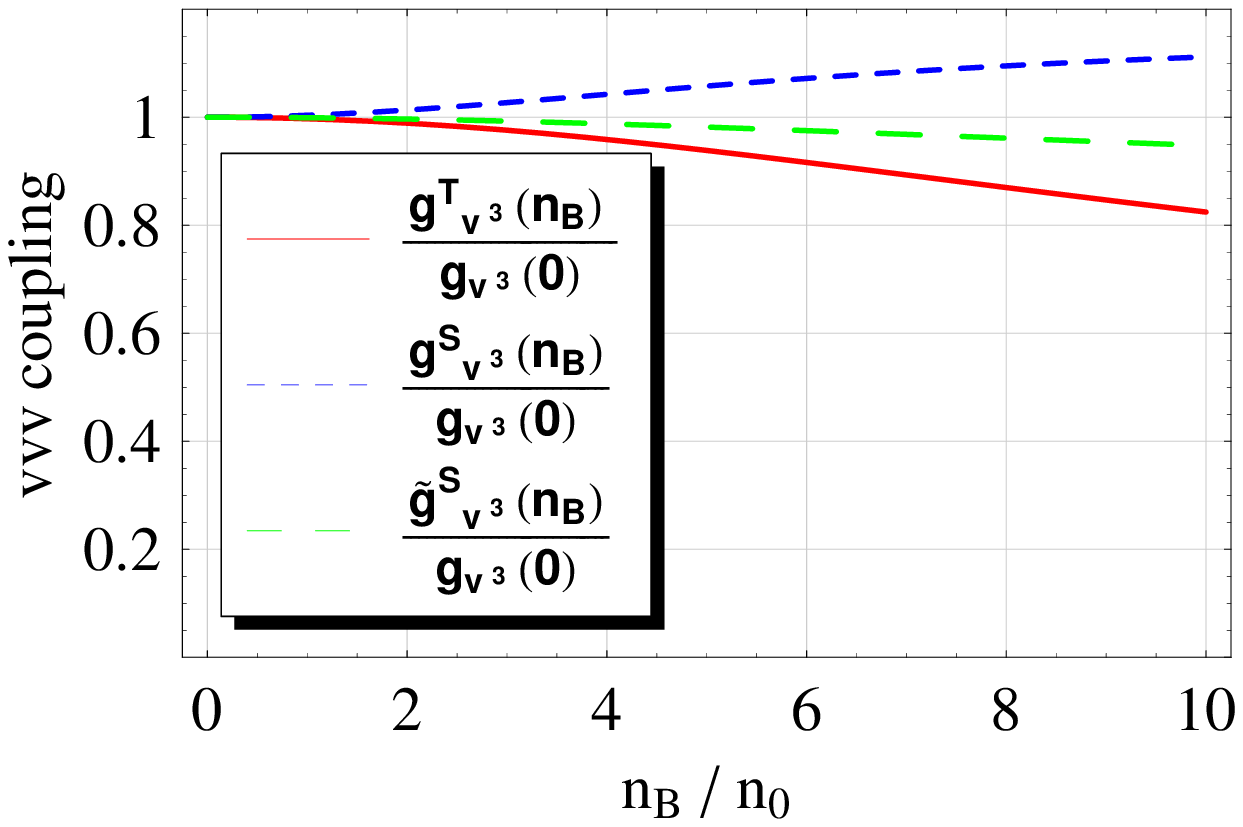}}
  \caption{(a) $v\Pi\Pi$ coupling (b) $vvv$ coupling }
  \label{Fig:KSRF}
  \end{center}
\end{figure}
\subsubsection{KSRF relations}
In the matter rest frame Lorentz symmetry is no longer manifest.
As a result, we expect a {\it variety} of KSFR relations
depending on wether time-like or space-like parameters are used.
Indeed, for instance the a-parameter at the origin of the KSFR
relations can now take 4 different forms depending on the
time-like/space-like arrangement. Specifically
\begin{eqnarray}
a_\textrm{KSRF}^{T1} \equiv \frac{4\, (g_{v\Pi^2}^T)^2\,
{(f_\pi^T)^2}}{m_1^2 } \ , \qquad a_\textrm{KSRF}^{T2} \equiv
\frac{4\, (g_{v\Pi^2}^T)^2\,
{(f_\pi^T)^2}}{ (\calm_{1}^\shortparallel)^2 } \nn \\
a_\textrm{KSRF}^{S1} \equiv \frac{4\, (g_{v\Pi^2}^S)^2\,
{(f_\pi^S)^2}}{(\calm_{1}^\perp)^2 } \ , \qquad
a_\textrm{KSRF}^{S2} \equiv \frac{4\, (g_{v\Pi^2}^S)^2\,
{(f_\pi^S)^2}}{ (\calm_{1}^\shortparallel)^2 } \nn \\
\end{eqnarray}
\begin{figure}[!hb]
  \begin{center}
   \includegraphics[width=8cm]{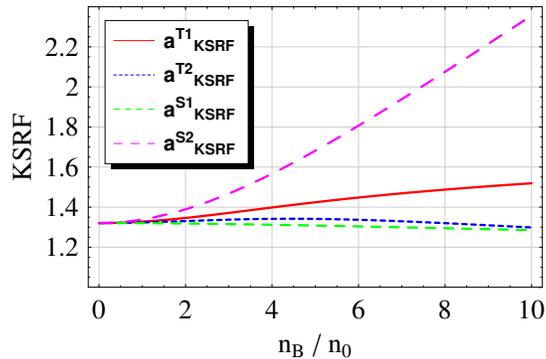}
  \caption{Generalized a-parameter}
  \label{fpi2tsscreening}
  \end{center}
\end{figure}

\section{Conclusions}

We have considered a generalization of the chiral model proposed
by Sakai and Sugimoto to finite baryon density. The baryon
vertices in bulk are attached equally to the D8-$\DeB$ branes and
correspond to $S^4$ in $D8$. They are treated as stable and point
like in $\mathbb{R}^3$ and act as uniform sources of baryon
density. Their point-like nature at large $N_c$ and coupling $\l$
imply that their interactions as induced by D8-$\DeB$ is mostly
repulsive through the exchanges of omega mesons.

The bulk energy density grows quadratically with the baryon
density before softening at asymptotic densities. The quadratic
and repulsive growth is expected from the exchange of omega
mesons. The softening reflects on the fact that at asymptotic
densities the repulsive baryons form an instable but regular
array for fixed volume $V$. If $V$ acting as a container is
removed, the baryons fly away in this version of the SS model. We
note that the energy density scales as $N_c$ since $N_c/\sqrt{a}$
is of order 1 as expected from standard large $N_c$ arguments.
The DBI action resums (partially) the strong NN-interactions while
keeping the leading $N_c$ result unchanged. Since the instanton
size is of order $1/\sqrt{\lambda}$ we also note that the resummed
contributions are of order $\lambda^0$ since the bulk instanton
density $\sqrt{\l}n_B$ is of
order $\lambda^{2}$ (The additional $\sqrt{\l}$ here stems from the
rescaling of $z \ra z/\sqrt{\l}$ in the delta-function source at $z=0$).

Using linear response theory, we have probed this dense baryonic
system using pions, vectors and axials. The point like nature of
the baryons with a size of order $1/\sqrt{\lambda}$ and the large
$N_c$ nature as noted above, causes rather mild changes in the
masses and couplings as a function of baryon density. In
contrast, the pion decay constants are found to change
appreciably. The quadratic increases at small baryon densities is
mediated by omega's. The scalar S-wave pion-baryon scattering
length is noted to vanish at large $N_c$, causing $f_\pi$ to
increase instead of decreasing at finite density. This behaviour
is unphysical.

The current approach needs to be improved in a number of ways to
accomodate the baryon physics expected in the real world. First,
the point-like nature of the sources need to be relaxed. This is
possible by constructing the pertinent instanton vertex. Also,
the point-like limit suggests that the DBI results quoted here are
only indicative since higher derivative corrections to the DBI
effective action are expected to contribute (see also
\cite{Sakai1,Sakai2,Sakai3} for further comments on this point)).
Second, the Fermi motion of the sources need to be included. This
can be achieved through a select quantization of the collective
variables associated to the baryon vertex insertion. Some of
these issues
will be addressed in later work.\\

\noindent {\bf Note added.} After the completion of this work, we
became aware of the recent work by O. Bergman, G. Lifschytz, and
M. Lippert~\cite{Bergman} who also address the SS model at finite
baryon density. They have shown that a cusp configuration develops
at finite density for generically separated D8-$\DeB$.  This
observation does not apply to the original SS embedding we
discuss here. We also noticed the appearance of two relevant
papers: \cite{Kraus} discusses the finite density problem in the
holographic NJL model, and \cite{Rozali} discusses the effects of
a finite size baryon charge distribution.

\section{Acknowledgments}
We would like to thank  Gerry Brown,  Shin Nakamura, Yunseok Seo,
 Edward Shuryak,  Patta  Yogendran and especially  Mannque Rho
 for useful  discussions on the subject of dense matter in AdS/CFT.
The work of KYK and IZ was supported in part by US-DOE grants
DE-FG02-88ER40388 and DE-FG03-97ER4014. The work of SJS was
supported by KOSEF Grant R01-2004-000-10520-0 and the SRC Program
of the KOSEF through the CQUEST with grant number R11-2005-021.

\appendix

\section{The Effective Action Using the Vacuum Modes}
\label{Pirhoaction}
Let us only consider the lightest vacuum meson modes corresponding
to the pion and $\rho$ meson fields.  This vacuum mode
decomposition was studied in~\cite{Sakai1} at zero baryon
density. Here we just add $\Az$ as obtained in
section~\ref{Sec:Chemical} to the gauge field $A_M$. Since the
mode decomposition is complete, this approach should be
complementary to the one discussed in the text. It is the same as
the one we used in~\cite{KSZ}. As we will show, the results are
overall similar to the ones discussed in the main text regarding
the density dependence.

In the gauge $A_z = 0$ and $\xi \equiv
e^{\frac{i\Pi(x^\m)}{f_\pi}}$ (\ref{AzZeroGauge}), $A_\m$ reads\footnote{In this section the
gauge field $A_\m$ is treated as anti-Hermitian. $\Az$ and $\Pi$
is Hermitian so $i$ was introduced, while $v_\m$ is
anti-Hermitian. Note that we are working in a different
gauge from Section~\ref{Newway}.}
\begin{eqnarray}
A_\mu(x^\mu,z)&=& - i\Az(z)+ v_\mu(x^\mu)\, \psi_1(z) \ . \nn \\
&+& \left(\frac{2i}{f_\pi}\dell_\mu \Pi + [\dell_\mu
\Pi^3]\right) \wh{\psi}_0(z) + \frac{1}{2 f_\pi^2}[\Pi, \dell_\mu
\Pi]  + \calo(\Pi^4) \ , \label{Amu}
\end{eqnarray}
where $\Az$ is the background field, $v_\m \equiv B_\mu^{(1)}$.
We have set $ B_\mu^{(n)}= 0 $ for $n \geq 2$. The corresponding
field strengths are
\begin{eqnarray}
F_{\mu\nu} &=&  (\dell_\mu v_\nu - \dell_\nu v_\mu) \psi_1 +
[v_\mu , v_\nu] \psi_i^2 \nn \\
&+& \frac{2i}{f_\pi}([\dell_\mu \Pi , v_\nu ]+[v_\nu , \dell_\nu
\Pi])\psi_1 \wh{\psi}_0 + \frac{1}{f_\pi^2}[\dell_\mu \Pi,
\dell_\nu\Pi](1-4\wh{\psi}_0^2) + \calo((\Pi, v_\mu)^3)  \\
F_{z\mu} &=& -i \dAz + \left(\frac{2i}{f_\pi}\dell_\mu \Pi +
[[\dell_\mu \Pi^3]]\right) \wh{\phi}_0 + v_\mu \dot{\psi}_1 +
\calo(\Pi^4)
\end{eqnarray}
where  $\dAz = \frac{d\Az}{dz} $, $\dot{\psi}_1 =\frac{d
\psi_1}{dz} $, and
\begin{eqnarray}
\wh{\phi}_0 =\dell_z \wh{\psi}_0 = \inv{\pi \Ukk}\inv{K} \sim
\phi_0 \textrm{ in (\ref{phi.0})}
\end{eqnarray}
Notice that $\Az$ does not contribute to $F_{\mu \nu}$ and affect
only $F_{z \mu}$.

In order to compute the DBI action (\ref{DBI.1}),
\begin{eqnarray}
&& S_{\DeDeB}^{DBI}
= -\wt T \int d^4x dz\, U^2  \nn \\
&&~~~~~~~~ \Tr \, \sqrt {\, 1 - (2\pi\alpha')^2 \frac{R^3}{2U^3}
 F_{\mu \nu}F^{\mu \nu}
- (2\pi\alpha')^2 \frac{9}{4}\frac{U}{\Ukk}  F_{\mu z}F^{\mu z} +
[F^3] + [F^4] + [F^5]  } \nn \ ,
\end{eqnarray}
we need to know $ F_{\mu \n}F^{\mu \n}, F_{\mu z}F^{\mu z}, [F^3],
[F^4]$, and $[F^5]$, which have many complicated contributions.
Again, we use the observations noted in the text to simplify. Thus
\begin{eqnarray}
[F^4] = (2\pi\alpha')^4
\frac{9}{8}\frac{U}{\Ukk}\left(\frac{R}{U}\right)^3
F_{0z}F^{0z}F_{ij}F^{ij} + \calo((v_\m,\varphi)^4) \ .
\end{eqnarray}
Table (\ref{t1}) lists all relevant terms. We have introduced
$f_{\m\n}$ defined as
\begin{eqnarray}
f_{\m\n} \equiv \dell_\m v_\n - \dell_\n v_\m \ ,
\end{eqnarray}
with $\mu\nu = 0,1,2,3$ and $i,j = 1,2,3$. Table (\ref{t1})
should be understood in the integral and trace operation. We have
omitted some terms vanishing in the operation and rearranged some
terms by using the cyclicity of the trace.
\begin{table}[]
\begin{center}
\begin{tabular}{|c c l c c|}
\hline \vspace{-0.2cm}
& & & &  \\
\vspace{-0.2cm}
$F_{\mu \nu}F^{\mu \nu}$ & $\ra$ & $  f_{\m\n}f^{\m\n}  \psi_1^2 $ & $\equiv$ & $\a_2$ \\
 & & & &  \\
$ $ & & $2 f_{\mu\nu} [v^\m, v^\n] \psi_1^3 + \frac{2}{f_\pi^2}f_{\m\n}[\dell^\m \Pi , \dell^\n \Pi]\psi_1(1-4\wh{\psi}_0^2) $ & $\equiv$ & $\a_3$\\
\vspace{-0.2cm}
& & & &  \\
\hline \vspace{-0.2cm}
& & & &  \\
$F_{\mu z}F^{\mu z} $ &$\ra$ & $ (\dAz)^2  $ & $\equiv$ & $\b_0 $\\
\vspace{-0.2cm} & & & &  \\
$ $ & & $ -\frac{4}{f_\pi}(\dell_0 \Pi)\wh{\phi}_0 \dAz  + 2i v_0 \dot{\psi}_1\dAz $ & $\equiv$ & $\b_1$ \\
\vspace{-0.2cm}& & & &  \\
$ $ & & $ -\frac{4}{f_\pi^2} (\del{\mu}\Pi \dmuup\Pi)
\wh{\phi}_0^2+
v_\m v^\m \dot{\psi}_1^2 + \frac{2i}{f_\pi} \{\dell_\m \Pi , v^\m \} \wh{\phi}_0 \dot{\psi}_1 $ & $\equiv$ & $\b_2 $ \\
\vspace{-0.2cm}& & & &  \\
\hline %
\vspace{-0.2cm}& & & &  \\
$ [F^4] $ &$\ra$ & $ f_{ij}f^{ij}(\dAz)^2\psi_1^2 $
& $\equiv$ & $ \g_2 $\\
\vspace{-0.2cm}& & & &  \\
$  $  & & $ \Big[ 2f_{ij}[v^i, v^j ](\dAz)^2 \psi_1^3 +
\frac{2}{f_\pi^2}f_{ij}[\dell^i \Pi , \dell^j
\Pi](\dAz)^2\psi_1(1-4\wh{\psi}_0^2) $ &
$ $ & $ $ \\
$  $  & &  $  -2i v_0 f_{ij}f^{ij}\dAz\dot{\psi}_1 \psi_1^2 \Big]
$
& $\equiv$ & $\g_3 $\\
 & & & &  \\
\hline
\end{tabular}
\end{center}
  \caption{The relevant terms in evaluating DBI action up to third order in the fields ($\Pi, v$).
           All entries are understood in the integral and trace operation. }
  \label{t1}
\end{table}

In terms of the entries in the RHS of the table, the
action reads
\begin{eqnarray}
 S_{\DeDeB}^{DBI} = -\wt T \int d^4x dz\, U^2 \, \Tr \, \sqrt {\, P_0 +
P_1} \ ,
\end{eqnarray}
with
\begin{eqnarray}
P_0 &\equiv& 1 - (2\pi\alpha')^2 \frac{9}{4}\frac{U}{\Ukk} \ \b_0
= 1 -
  b K^{\frac{1}{3}}(\dZAz)^2  \ , \\
P_1 &\equiv&   (2\pi\alpha')^2 \frac{ R^3}{2 U^3}(\a_2 + \a_3) +
(2\pi\alpha')^2 \frac{9}{4}\frac{ U}{\Ukk} (\b_1+ \b_2 ) \nn \\
&+&(2\pi\alpha')^4 \frac{9}{8} \frac{ R^3}{U_{KK}U^2 } (\g_2 +
\g_3) \ . \label{P1}
\end{eqnarray}
Again, $P_0$ does not include meson fields and has carries the
baryon density. Expanding the action by fluctuating the fields we
have
\begin{eqnarray}
S_{\DeDeB}^{DBI} &=& -\wt T \int d^4x dz\, U^2 \,\Tr \, \left[
\sqrt {\, P_0} + \half \frac{P_1}{\sqrt{P_0}}
-\frac{1}{8}\frac{P_1^2}{{\sqrt{P_0}}^3} +
\frac{1}{16} \frac{P_1^3}{\sqrt{P_0}^5} \right] + \calo((\Pi, v_\mu)^4) \nn \\
&=& S_1 + S_2 +  \calo((\Pi, v_\mu)^4) \ ,
\end{eqnarray}
with
\begin{eqnarray}
&& S_1 \equiv -\wt T \int d^4x dz\, U^2 \Tr \, \D^{-1} \ , \\
&& S_2 \equiv -\wt T \int d^4x dz\, U^2 \Tr \, \left[\half \D P_1
 - \frac{1}{8} \D^3 P_1^2  +
\frac{1}{16} \D^5 P_1^3 \right]\  ,  \label{S2}
\end{eqnarray}
where we defined a modification factor $\D(Q)$ as
\begin{eqnarray}
\D(Q) \equiv  \frac{1}{\sqrt{P_0}} = \frac{1}{\sqrt {\, 1 -
  b K^{\frac{1}{3}}(\dZAz)^2} } = \sqrt{1+\frac{n_B^2} {4 a^2 b} K^{-5/3} } \ . \nn
\end{eqnarray}
Notice that $-S_1$ is the grand potential discussed in section
\ref{Sec:Chemical}, and $S_2$ will be reduced to the action of
mesons. To accomplish it we plug (\ref{P1}) into (\ref{S2}) and
evaluate all $z$ integrals and identify them as coefficients of
each term in the remaining 4-D action.

Let us first check which terms we have and how they are affected
by finite baryon density schematically. It can be read off from
Table (\ref{t1}). At zero density we set $\Az=0$ and $\D = 1$.
Then $\a_2, \a_3, \b_2$ survive. $\a_2,\b_2$ correspond to the free
action of $\Pi$ and $\rho$, and $\a_3$ is the couplings of $vvv$,
$v\Pi\Pi$ interaction. At finite density all terms are enhanced by
$\D,\D^2,$or $\D^3$. Furthermore there are nontrivial
modification. The free action part will be affected by $\g_2$ and
$\b_1^2$. ($\b_1$ itself does not contribute because the first
term has odd parity in $z$ and the second term is traceless.) The
couplings are modified by $\g_3$. There are new interaction terms
such as $ \bvo (\dell_\mu v_\nu-\dell_\nu v_\mu )^2,\ v_0 v_\mu
v^\mu,\ v_0 \dell_\mu \Pi \dell^\mu \Pi,\ \dell_0 \Pi \{\dell_\mu
\Pi, v^\mu\}$, which all vanish at zero density.

\begin{table}[!t]
\begin{center}
\begin{tabular}{|c|l|c|}
\hline
\vspace{-0.2cm} & &  \\
Coefficients & \hspace{2cm} Definition & $Q=0$($\Delta=1)$ \\
\vspace{-0.2cm}& &   \\
\hline \hline
\vspace{-0.2cm}& &  \\
$a_{\Pi^2}^T$ & $\frac{1}{\pi}\int dZ K^{-1} \,
\D^3  $ & $1$ \\
\vspace{-0.2cm}& &  \\
$a_{\Pi^2}^S$ & $\frac{1}{\pi}\int dZK^{-1} \, \Delta$ & $1$\\
\vspace{-0.2cm}& &  \\
\hline
\vspace{-0.2cm}& &  \\
$a_{v^2}^T$ & $\int dZ\,  K^{-\frac{1}{3}} \Psi_1^2 \, \Delta $ & $1$\\
\vspace{-0.2cm}& &  \\
$a_{v^2}^S$ & $\int dZ\,  K^{-\frac{1}{3}} \Psi_1^2 \, \Delta^{-1}  $& $1$\\
\vspace{-0.2cm}& &  \\
\hline
\vspace{-0.2cm}& &  \\
${m_v^2}^T$ & $\Mkk^2 \int dZ\, K
(\dell_Z\Psi_1)^2 \, \D^3  $ & $m_\rho^2$ \\
\vspace{-0.2cm}& &  \\
${m_v^2}^S$ & $\Mkk^2
\int dZ\, K (\dell_Z\Psi_1)^2 \, \Delta $ & $m_\rho^2$ \\
\vspace{-0.2cm}& &  \\
\hline & &  \\
$a_{v^3}^T$  & $\inv{\sqrt{\k}}\,\int
dZ\,K^{-\frac{1}{3}} \Psi_1^3 \, \Delta $
& $\inv{\sqrt{\k}} \cdot 0.446 $\\
\vspace{-0.2cm}& &  \\
$a_{v^3}^S$  & $\inv{\sqrt{\k}}\,\int dZ\, K^{-\frac{1}{3}}
\Psi_1^3 \, \Delta^{-1}  $
& $\inv{\sqrt{\k}} \cdot 0.446 $\\
\vspace{-0.2cm}& &  \\
\hline
\vspace{-0.2cm}& &  \\
$a_{v\Pi^2}^T$ & $\inv{\sqrt{\k}}\frac{\pi}{4 \Mkk^2} \int
dZ\, K^{-1/3} \Psi_1 (1-4\wh{\psi}_0^2) \D  $ & $\inv{\sqrt{\k}}\frac{\pi}{4 \Mkk^2}\cdot 1.584 $\\
\vspace{-0.2cm}& &  \\
$a_{v\Pi^2}^S$ & $\inv{\sqrt{\k}}\frac{\pi}{4 \Mkk^2} \int
dZ\, K^{-1/3} \Psi_1 (1-4\wh{\psi}_0^2) \D^{-1} $ & $\inv{\sqrt{\k}}\frac{\pi}{4 \Mkk^2}\cdot 1.584 $ \\
 & &  \\
\hline
\end{tabular}
\end{center}
   \caption{The definitions of the coefficients in the action (\ref{S2final}).
            At finite density, there are enhancing factors $\D, \D^3$ and a suppressing factor $\D^{-1}$. }
    \label{t2}
\end{table}
Considering all these modification we get the final form of the
meson action
\begin{eqnarray}
S_2 = \int d^4 x \Bigg[ &-& a_{\Pi^2}^T
\Tr\left(\dell_0\Pi\dell^0\Pi\right) -a_{\Pi^2}^S
\Tr\left(\dell_i\Pi\dell^i\Pi\right)\nn \\
&+& a_{v^2}^T\Tr f_{0i}f^{0i} +
\frac{1}{2}a_{v^2}^S\Tr f_{ij}f^{ij} \nn \\
&+& {m_v^2}^T \Tr v_0 v^0 + {m_v^2}^S\Tr v_iv^i\nn\\
&+& a_{v^3}^T \Tr\Big( 2 f_{0i}[v^0,v^i] \Big) +a_{v^3}^S \Tr\Big(  f_{ij}[v^i,v^j] \Big) \nn \\
&+& a_{v\Pi^2}^T \Tr\Big(2f_{0i}[\dell^0 \Pi, \dell^i \Pi]\Big)
+a_{v\Pi^2}^S
 \Tr\Big(f_{ij}[\dell^i \Pi, \dell^j
\Pi]\Big) + \cdots \Bigg] \ , \label{S2final}
\end{eqnarray}
where the coefficients of every term are defined in Table
(\ref{t2}). At zero density all coefficients agree with those
in~\cite{Sakai1}. There are three types of modification due to
the baryon density: $\D ,
\D^{-1},$ and $\D^{3}$. $\D$ simply comes from  $\half \D P_1 $ in
(\ref{S2}). Since $\D$ is common to all coefficients, $\D^{3}$ and
$\D^{-1}$ can be understood as $\D \cdot \D^2$ and $\D \cdot
\D^{-2}$. An enhancing factor $\D^2$ is due to additional
contribution from $\b_1^2$ and a suppressing factor $\D^{-1}$ is
from $\g_2, \g_3$, which explains the calculational similarities
in all the results.

Finally we want to mention  without details, the character of the
action in the gauge $A_M(z \ra \infty) \ra 0$ instead of $A_z = 0$.
The action is the same as in~(\ref{S2final}) except for the interaction
term $v \Pi\Pi $
\begin{eqnarray}
-2g^T_{v\Pi^2}\, v_0 [\Pi , \dell^0 \Pi ] -2g^S_{v\Pi^2}\, v_i
[\Pi , \dell^i \Pi ] \ ,
\end{eqnarray}
where $g^T_{v\Pi^2}$ and $g^S_{v\Pi^2}$ are defined in
Table(\ref{t4}).
\begin{table}[!t]
\begin{center}
\begin{tabular}{|c|l|c|}
\hline
\vspace{-0.2cm} & &  \\
Coefficients & \hspace{2cm} Definition & $Q=0$($\Delta=1)$ \\
\vspace{-0.2cm}& &   \\
\hline \hline
\vspace{-0.2cm}& &  \\
$g_{v\Pi^2}^T$ & $\sqrt{\k}\Mkk^2 \Ukk^2 \int
dZ\, K \D^{3} \Psi_1 \phi_0^2 $ & $\inv{\sqrt{\k} \pi}\cdot 0.63 $\\
\vspace{-0.2cm}& &  \\
$g_{v\Pi^2}^S$ & $\sqrt{\k}\Mkk^2 \Ukk^2 \int
dZ\, K \D \Psi_1 \phi_0^2 $ & $\inv{\sqrt{\k} \pi}\cdot 0.63 $ \\
 & &  \\
\hline
\end{tabular}
\end{center}
   \caption{The definitions of the coefficients in the action (\ref{S2final}).
            At finite density, there are enhancing factors $\D, \D^3$ and a suppressing factor $\D^{-1}$. }
    \label{t4}
\end{table}

\subsection{Numerical results}
In this section we compute the coefficients in Table
(\ref{t2},\ref{t4}) numerically. Their physical meanings can be
read off from the action (\ref{S2final}). We use the same
numerical inputs as detailed in section 3.
\subsubsection{Pion Decay Constant}
The pion decay constant can be defined by the procedure of Section
\ref{Skyrme.Action} with the vacuum mode function.

\begin{eqnarray}
f_\pi^{T} \equiv f_\pi \sqrt{ a_{\pi^2}^T}  \ ,\quad \quad
f_\pi^{S} \equiv f_\pi \sqrt{ a_{\pi^2}^S}
\end{eqnarray}
\subsubsection{Velocity}
The pion velocity:
\begin{eqnarray} \label{pivelocity}
v_\pi \equiv \sqrt{\frac{a_{\pi^2}^S}{a_{\pi^2}^T}} =
\frac{f_\pi^{S}}{f_\pi^{T}}
\end{eqnarray}
The lowest mode velocity:
\begin{eqnarray}
v_v \equiv  \sqrt{\frac{a_{v^2}^S}{a_{v^2}^T}}
\end{eqnarray}
\subsubsection{Mass}
\begin{eqnarray}
M_1 \equiv \sqrt{\frac{m_v^{2S}}{a_{v^2}^T}}
\end{eqnarray}
\subsubsection{Screening mass}
\begin{eqnarray}
M_{scr}^{\shortparallel} \equiv \sqrt{\frac{m_v^{2T}}{a_{v^2}^T}}
\ ,\quad \quad M_{scr}^{\perp} \equiv
\sqrt{\frac{m_v^{2S}}{a_{v^2}^S}} \ .
\end{eqnarray}
\begin{figure}[]
  \begin{center}
  \subfigure[] {\includegraphics[width=8cm]{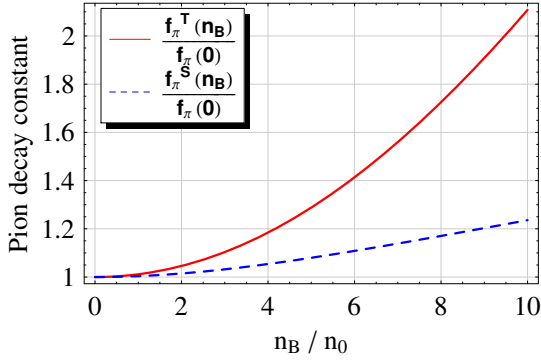}}
  \subfigure[] {\includegraphics[width=8cm]{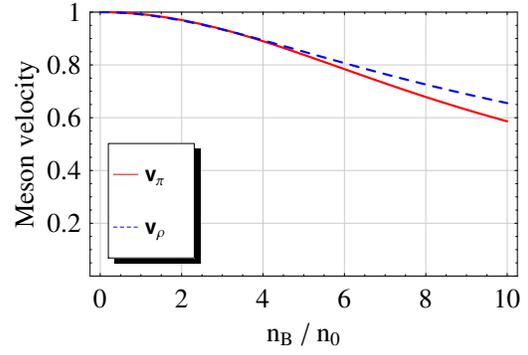}}
  \caption{(a) Pion decay constant vs $\frac{n_B}{n_0}$ \big[$\frac{f_\pi^T}{f_\pi} = \sqrt{ a_{\pi^2}^T} $,
          $\frac{f_\pi^{S}}{f_\pi} = \sqrt{ a_{\pi^2}^S}$ \big],
           (b) Velocity of $\Pi$ and $\rho$ (v) vs $\frac{n_B}{n_0}$
  \big[
     $v_\pi \equiv \sqrt{\frac{a_{\pi^2}^S}{a_{\pi^2}^T}}$ and
     $v_v \equiv  \sqrt{\frac{a_{v^2}^S}{a_{v^2}^T}}$ \big]}
  \label{fpi2tsvelo}
  \end{center}
\end{figure}
\begin{figure}[]
  \begin{center}
  \subfigure[] {\includegraphics[width=8cm]{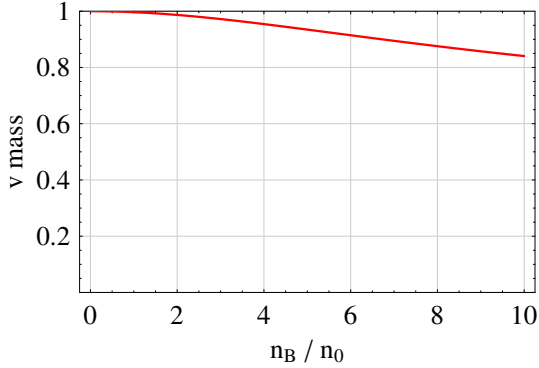}}
  \subfigure[] {\includegraphics[width=8cm]{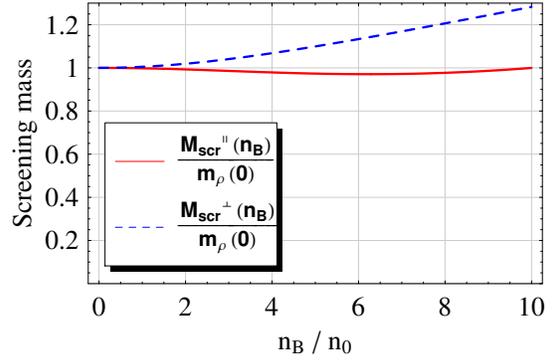}}
  \caption{(a) $v$ mass vs $\frac{n_B}{n_0}$ \big[$\frac{M_1}{m_\rho} \equiv \sqrt{\frac{m_v^{2S}}{a_{v^2}^T m_\rho^2}}$ \big] ,
        (b) Screening masses vs $\frac{n_B}{n_0}$ $\big[
      \frac{M_{scr}^{\shortparallel}}{m_\rho} \equiv \sqrt{\frac{m_v^{2T}}{a_{v^2}^T m^2_\rho}}$
      , $\frac{M_{scr}^{\perp}}{m_\rho} \equiv
\sqrt{\frac{m_v^{2S}}{a_{v^2}^S m_\rho^2}} \big]$ }
  \label{fpi2tsvelo}
  \end{center}
\end{figure}

\end{document}